\documentclass[a4paper,11pt]{article}
\usepackage[mathscr]{eucal}
\usepackage{amsmath,amsfonts,amssymb,amsthm}
\usepackage{times}
\usepackage{hyperref}

\newcommand{\maxim}[1]{\textsc{#1}}

\advance\textheight\topskip
\numberwithin{equation}{section} \makeatletter
\@addtoreset{equation}{section}

\newtheorem{prop}{Proposition}[section]

\renewcommand{\tilde}{\widetilde}
\renewcommand{\hat}{\widehat}

\newcommand{\bref}[1]{\textbf{\ref{#1}}}

\newcommand{\p}[1]{|#1|}

\newcommand{\dd}{\partial}
\renewcommand{\d}{\partial}

\renewcommand{\geq}{\,{\geqslant}\,}
\renewcommand{\leq}{\,{\leqslant}\,}

\newcommand{\inner}[2]{\langle #1{,}\,#2\rangle}
\newcommand{\binner}[2]{%
  {\langle}\kern-4.15pt{\langle}#1{,}\,#2{\rangle}\kern-4.15pt{\rangle}}
\newcommand{\commut}[2]{[#1{,}\,#2]}
\newcommand{\scommut}[2]{\{#1{,}\,#2\}}
\newcommand{\sqcommut}[2]{\{#1{,}\,#2\}_*}
\newcommand{\qcommut}[2]{[#1{,}\,#2]_*}
\newcommand{\pb}[2]{\left\{{}#1{},{}#2{}\right\}}

\newcommand{\half}{\mathchoice{%
    \ffrac{1}{2}}{\frac{1}{2}}{\frac{1}{2}}{\frac{1}{2}}}

\newcommand{\ffrac}[2]{\raisebox{.5pt}%
  {\footnotesize$\displaystyle\frac{#1}{#2}$}\kern1pt}

\newcommand{\brst}{\mathsf{\Omega}}

\newcommand{\dl}[1]{\mathchoice{\ffrac{\dd}{\dd #1}}{\frac{\dd}{\dd
      #1}}{\ffrac{\dd}{\dd #1}}{\ffrac{\dd}{\dd #1}}}
\newcommand{\dr}[1]{\ffrac{{\overset{\leftarrow}{\partial}}}{ \partial #1}}

\newcommand{\vddl}[2]{{\ffrac{\delta #1}{\delta #2}}}

\newcommand\Rho{\mathfrak{R}}

\def\cA{\mathcal{A}}

\def\cD{\mathcal{D}}

\def\cR{\mathcal{R}}

\def\cT{\mathcal{T}}

\def\tr{{\rm Tr}}

\def\BG-Poincare{Barnich:2009jy}
\def\Fedosov-book{Fedosov:1996fu}

\usepackage{ulem}

\begin{document}


\begin{flushright}\small{Imperial-TP-AT-2016-{04}}\end{flushright}
\vspace{1.5cm}
\begin{center}
 \textbf{\Large
On conformal higher spins in curved  background  }

\vspace{1.0 cm}

  {\large {M. Grigoriev$^a$  and A.A. Tseytlin$^{a,b}$}}
\vspace{.5cm}
  %
%

$^a$\textit{Tamm Theory Department,
 Lebedev Physics Institute\\
 Leninsky prospect 53, 119991 Moscow, Russia}
 
 \vspace{0.2cm}
 
$^b$\textit{Theoretical Physics Group, Blackett Laboratory,\\ Imperial College London,   SW7 2AZ, U.K.
}
\end{center}
\vspace{1cm}

\begin{abstract}
We   address the question of 
 how to  represent an  interacting  action for a tower of conformal  higher spin  fields 
in a form covariant with respect to a  background metric. We use the background metric to  define 
the  star product which plays a central role in the definition of the corresponding  gauge transformations. 
By analogy with  the  kinetic term  in the  4-derivative Weyl  gravity action expanded 
near an on-shell background    one expects that the kinetic term in
 such an action   should be gauge-invariant  in a Bach-flat  metric. 
 We demonstrate this fact to first   order  in expansion in powers of  the curvature  of the background metric.
 This  generalizes   the result of   arXiv:1404.7452 for spin 3 case to all conformal higher spins. 
We also  comment  on a possibility of extending  this claim to terms quadratic in the curvature 
and discuss the appearance of  background-dependent mixing terms in the 
 quadratic part of the conformal higher spin action. 
\end{abstract}

\thispagestyle{empty}

\vspace{1cm}
\newpage 
\setcounter{tocdepth}{2}
\tableofcontents
\def \la {\label}
\def \fo {{\textstyle { 1 \ov 4}}}
\def \la {\label}
\newcommand \foot [1] {\footnote{#1\vspace{2pt}}}
\newcommand \rf [1] {(\ref{#1})}
\newcommand \sfrac [2] {{\textstyle\frac{#1}{#2}}}
\newcommand \vev [1] {\langle{#1}\rangle}
\def \str {{\rm str}}
\def \Tr {{\rm Tr}}
\def \ha {{ \textstyle{1 \over 2}}}
\def \td {\tilde}
\def \ci{\cite}
\def \del{ \partial}
\def\ov{\over}
\def\no{\nonumber} 
\def \aa {{\rm a}}
\def \p {\phi}
\def \m {\mu}\def \n {\nu} 
\def \te {\textstyle} 
\def \ed  {\end{document} }
\def \fo {{ \textstyle{1 \over 4}}}
\def \be {\begin{equation}}
\def \ee {\end{equation}}
\def \ed {\end{document}}
\def \ba {\begin{align}}
\def \era {\end{align}}
\def \PB {{\rm PB}}

\renewcommand{\r}{\mathbf{r}}
\renewcommand{\d}{\partial}

\def \OO {{\cal J}} 
\def \lR {L}
\def \ll {{\cal \ell}}
\def \be {\begin{equation}}\def \ee {\end{equation}}
\def \ads {AdS$_4$\ }
\def \iffa {\iffalse} \def \G {\Gamma} 
\def \SS {{\rm S}}
\def \ed {\end{document}}
\def \k {\kappa} 

\def \te {\textstyle} 

\section{Introduction}

Conformal higher spin  (CHS)  field models   are $s > 2$ generalizations   of  Maxwell ($ F^2_{\m\n}$)   and  Weyl ($ C^2_{\m\n\lambda\rho}$) 
theories   \ci{Fradkin:1985am,Fradkin:1989md,Tseytlin:2002gz,Segal:2002gd}. 
While they   have  higher-derivative $\del^{2s}$   kinetic terms  and thus are formally non-unitary they  have a remarkable feature  of  describing 
pure spin $s$ states  off-shell,  i.e. have  maximal  spin $s$ gauge symmetry consistent with locality. 

The free CHS  action in flat  4-dimensional space  may be written as 
\be 
S_s= \int d^4x \  h_s  P_s \, \del^{2s}\,  h_s  = \int d^4 x\    (-1)^s \, C_{s}  C_{s}
  \ , \la{1} \ee
where $h_s= (h_{\m_1...\m_s})$ is a totally symmetric tensor and 
$P_s= (P^{\m_1...\m_s}_{\n_1...\n_s})$  is the transverse  projector  which is  traceless  
and  symmetric within  $\mu$ and $\n$ groups of indices.
This  action is thus invariant under a combination of 
differential (generalized reparametrizations)  and algebraic (generalized  Weyl)  gauge transformations: 
$\delta h_s = \del \epsilon_{s-1}  + \eta_2  \omega_{s-2}$  (here $\eta_2$ is a flat    metric and $\epsilon$ and $\omega$ are parameter tensors).
 $C_s
 = (C_{\m_1...\m_s,\n_1 ...\n_s})$  is  the gauge-invariant  field  strength or generalized Weyl tensor.

The  theory containing an   infinite tower of CHS fields $h_s$  ($s=0,1,2,...$)    is  a non-trivial interacting field theory 
with an  action   that    can be  defined as a local part of   an  induced  action \ci{Tseytlin:2002gz,Segal:2002gd,Bekaert:2010ky,Giombi:2013yva}.
Explicitly, one may start with a free CFT of $N$ scalar fields $\int d^4 x \  \phi^*_i \del^2 \phi_i $  which has 
the on-shell conserved   and traceless   spin $s$ currents  $J_s = \phi^*_i \OO_s  \phi_i$  \ ($  \OO_s \sim \del^s + ...$)
and  consider the  generating functional for correlation  functions of these currents 
\be\la{13}
	\G[h]= N \,\log \det  \Delta (h)  \ , \ \ \ \qquad  \   \Delta(h) =  -\del^2   + \sum_s h_s\,\OO_s \,.
\ee
 Here   $h_s(x)$   are source fields  which   have      linearized gauge symmetries  implied by   the on-shell   conservation and tracelessness of the currents $J_s$.\foot{According to vectorial AdS/CFT this induced action   should  follow  also from the  massless higher spin   theory in the $AdS_5$ bulk    upon  computing  it on the  solution  of the  equations of  motion  with $h_s$  setting the boundary conditions for the 5d massless    higher spin fields.} The UV logarithmically divergent  part  of 
  \rf{13}  is  local,  
  has the required  linearized gauge  symmetries and expanded in $h_s$
    starts with \rf{1}  as its  quadratic term.\foot{One  gets 
  $\G[h]=N\,\sum_s \int h_s  K_s  h_s  +   O(h^3) \ ,$ where 
  $K_s 
\sim   N^{-1}   <  J_s(x)   J_s(x') >\  \sim  P_s\,|x-x'|^{ -4 - 2s} 
\sim P_s\,\del^{2s }\,\delta^{(4)} (x-x')\, \log \Lambda +...$. 
Let us note that 
to get diagonal  kinetic terms  for all CHS fields one  needs to apply a certain  field redefinition 
required to make the algebraic Weyl-type  symmetry manifest \ci{Segal:2002gd}.}
  The  coefficient of the logarithmic divergence (or, equivalently, the $t^0$ Seeley   coefficient  in the small $t$ expansion of the 
  heat kernel of the  operator  $ \Delta (h)$)   can thus    be 
 taken as a definition of the  full  CHS  action, i.e.
\be\la{133}
	S[h]= N \, \big[\log \det  \Delta (h) \big]_{\log \Lambda} \sim  N\,  \tr\,  e^{-t  \Delta (h) }\Big|_{t=0} \ . 
\ee
In this particular construction  $N$ plays the role of the square of the inverse coupling  constant  which,  in general,  can be arbitrary. 
    A discussion of some  cubic  and quartic terms in this action  appeared in 
    \ci{Bekaert:2010ky,Joung:2015eny,Beccaria:2016syk}.\footnote{Since the dimension of $h_s$ is $2-s$ 
    and the  theory is scale invariant 
    the $h^m$ ($m=3,4,...$)  interaction   vertex   containing fields of  spins $s_i$ ($i=1,\ldots,m$)
involves   $ k= 4 +\sum_{i=1}^m\,(s_i-2) $  derivatives. Thus the  coupling to the dimension 0  spin 2  field (conformal graviton) 
is special: one may add an arbitrary number of $h_2$   factors in the vertex without increasing the number of derivatives.}

This  CHS theory     has  a close connection   to   AdS/CFT   but  has  also   several remarkable features on its own.
On general grounds, the  theory  $S_{\rm CHS}\sim  \int d^4 x ( h_0^2 + F_{\m\n}^2  + C^2_{\m\n\lambda\rho} + ....)$
with  dimensionless coupling  constant  should be renormalizable -- the gauge symmetries should 
fix the local  action uniquely. The central question is the absence of  anomalies, in particular, the Weyl  anomaly. 
  It was found  in \ci{Giombi:2013yva,Tseytlin:2013jya} 
that the one-loop  $a$-coefficient of Weyl anomaly of the $d=4$ CHS   theory  vanishes under 
a particular prescription   (which should be consistent with the underlying symmetries, see also 
  \ci{Giombi:2014iua,Beccaria:2015vaa}) for   summation over spins.
    The same   was  found  also    for the one-loop   conformal  anomaly 
  $c$-coefficient  \cite{Tseytlin:2013jya,Giombi:2014iua,Beccaria:2014xda,Beccaria:2015vaa}
  under the assumption 
  that contributions  to the  conformal anomaly from higher derivative   CHS  operators 
  on Ricci flat background  factorize.\foot{The  computation 
of the one-loop  conformal anomaly $c$-coefficient in the CHS theory in \ci{Tseytlin:2013jya} was  based on  two assumptions: 
(i)   the CHS action obtained as a UV divergent part of the  induced action  in near-flat space expansion can be 
reformulated (using  a field redefinition)  in such a way that  at least    quadratic kinetic terms 
 in generic  curved metric background  are     reparametrization and Weyl   covariant;
 (ii) the higher derivative kinetic   operators $\nabla^{2s} + ...$,  while not factorizing, in general, 
 into products of $\nabla^2+...$ operators on a Ricci-flat  background \ci{Nutma:2014pua}
 (as they do on    $AdS$ or on the  sphere) 
 still  contribute to the $c$-anomaly  in the same  way as  if they were  factorizing. The reason is  that   
 the  terms with derivatives of the curvature tensor that obstruct the  factorization can not contribute to  the $C^2_{\m\n\k\l}$ 
 conformal anomaly on dimensional grounds.}
As the Weyl   symmetry is one of the   CHS  gauge 
 symmetries, this is an indication that the same anomaly cancellation  may apply to all  algebraic  CHS symmetries.

  The CHS theory  has also the vanishing Casimir energy on $R \times S^3$ \ci{Beccaria:2014jxa} and 
   zero  total number of degrees of freedom 
(trivial flat-space partition function)  \ci{Beccaria:2015vaa}   which is a reflection of  the  large
underlying  gauge symmetry of this  theory. The global part of this    symmetry  also  strongly  constrains  the    S-matrix   involving exchanges of the CHS fields  implying that it should be trivial  \ci{Joung:2015eny,Beccaria:2016syk}. 

The action \rf{133}  is naturally defined as an expansion in powers of $h_s$  fields  near flat space. 
It can thus  be interpreted as a {higher spin  interacting  classical  conformal field theory}.
 One may then  wonder  if  it may admit a  reparametrization and Weyl covariant generalization  to a curved background 
  which is  known to exist  for the  standard  low-spin ($s=1,2$)  cases. 
 Assuming that  the $s=2$ field $h_{\m\n}$   may  be interpreted as  the   conformal graviton, 
one may ask  if the action \rf{133} can be rewritten (after some field redefinitions) 
 as an expansion   near  a curved background $g_{\m\n} \to g_{\m\n} + h_{\m\n}$. 

Here we  will restrict attention to terms in the CHS action that are quadratic in the  fields $h_s$ 
 but all-order in the background metric $g_{\m\n}  $  and address the question  which 
 background  geometries admit a consistent (gauge-covariant) propagation of $h_s$.  
 It follows from the flat-space conformal invariance that the CHS  field can be consistently defined   
  on any conformally flat background $g_{\mu\nu}(x)=\sigma(x)\eta_{\mu\nu}$. 
  In the case of  an arbitrary 
   $\sigma(x)$ the form of the   generic  spin $s$  kinetic operator  is not known explicitly 
 but can be reconstructed,  in principle,  by a $\sigma$-dependent rescaling of the field  (assuming there exists a Weyl-invariant generalization of the flat-space action \rf{1}).\foot{Alternatively,   including  some auxiliary and Stuekelberg fields one can reformulate
the  CHS action  in a manifestly conformal form for which rewriting in a  generic conformally-flat background amounts to just picking an appropriate 
$o(d,2)$-connection and  conformal compensator.}
 In the case of a homogeneous   conformally-flat space (4-sphere or AdS  or dS or $R \times S^3$) 
 the CHS  kinetic operator  is known   and  can be represented as a  product of second-order  differential operators 
 \ci{Tseytlin:2013jya,Metsaev:2007rw,Metsaev:2014iwa,Nutma:2014pua,Beccaria:2014jxa}. 
 The  question  is   whether the  CHS  fields  can be consistent on  non-conformally-flat backgrounds   
  with  non-vanishing Weyl tensor  and 
    what  are  the   conditions on the Weyl tensor for this to happen. 
 
 For   $s=1$ (Maxwell) and $s=2$ (Weyl)   cases the  CHS  kinetic terms admit   the well-known 
 generalizations to a non-trivial  background  metric $g_{\m\n}$. 
 For $s=1$  we get no constraints on  $g_{\m\n}$ while for $s=2$   the invariance of the   quadratic term in the 
 Weyl  action  $\int d^4 x \sqrt g\, C^2_{\m\n\lambda\rho}$   expanded  in $h_{\m\n}$ 
   (with $g_{\m\n} \to  g_{\m\n} + h_{\m\n}$)    gives a  special  $\nabla^4+...$    kinetic
 operator   \ci{Fradkin:1981iu,Fradkin:1985am,Beccaria:2014jxa}. This   operator 
 is  covariant under the  gauge transformations  
 $ \delta h_{\mu\nu}=\nabla_\mu \epsilon_\nu+\nabla_\nu \epsilon_\mu+\omega g_{\mu\nu}$ 
  provided $g_{\m\n}$  is an on-shell   background for
 the  Weyl theory, i.e.  is Bach-flat, 
\begin{equation}\la{55}\te B_{\mu\nu}=0\ , \qquad 
 B_{\mu\nu}\equiv \nabla^\rho\nabla_\mu P_{\nu\rho}-\nabla^\rho\nabla_\rho P_{\mu\nu} 
 - P^{\rho\sigma}C_{\mu\rho\nu\sigma}\,, \quad 
 P_{\mu\nu}\equiv\frac{1}{2}\big( R_{\mu\nu}-\frac{1}{6} Rg_{\mu\nu} \big) \ . 
 \end{equation} 
For $s =3$  CHS field 
this  question was addressed  in \ci{Nutma:2014pua} where the corresponding covariant $\nabla^6+...$ 
   kinetic operator   was found   to linear order in the   background curvature tensor  
    and was shown to be   gauge-invariant     on  Bach-flat backgrounds {(to first order in the curvature)}.
 
 A goal of the present paper  is    to make a step towards  a  covariant 
 description of  all CHS fields  on curved  Bach-flat  (or, in particular, Ricci-flat) backgrounds. 
Our starting point will be an equivalent definition of the non-linear CHS action \rf{133}   based on an effective particle Hamiltonian
associated with  the operator $\Delta(h)$ in \rf{13}  \ci{Segal:2002gd} that  makes the full non-linear symmetry of the theory 
more  explicit.

In section 2  we shall  review  the definition of the  particle   Hamiltonian  coupled to the 
CHS fields following   \cite{Segal:2002gd,Bekaert:2010ky}.  Its quantization  leads to a quadratic  scalar action  
in  CHS background  that has  gauge invariances  inherited from the  freedom in the definition of the particle dynamics. 
  In section 3   we shall  suggest a procedure of   how to define the scalar action in a way  covariant with respect to a background metric and having the required gauge symmetries.
   Then  the  corresponding CHS   action  can be again defined as a UV singular part 
   of the  induced action  found   after  integrating out the scalar field.
   
   In  section 4  we shall analyse the expansion of this action in powers of the CHS fields  and the  consistency 
   conditions of this  expansion using  perturbation theory in powers of the 
   curvature  of the background metric. 
   Section 5 will contain some  concluding remarks. 
In Appendix A  we shall  review the Fedosov-type  approach to covariant formulation of first-quantized particle dynamics
that plays important role in our definition of  the CHS   gauge transformations in a non-trivial background.  In Appendix B 
we shall make some general comments on the structure of Weyl invariants  built out of the curvature and its covariant derivatives.

\section{Particle Hamiltonian in CHS   background 
and   expansion near flat space}

Before developing a covariant approach to CHS fields let us briefly recall how their  
gauge transformations
and gauge-invariant action arise from the coordinate-dependent  quantized  particle  approach  \ci{Segal:2002gd,Segal:2001di}.

\subsection{Gauge transformations}

Let us  start  with a space-time manifold  
with coordinates $x^\mu$ and introduce   the momenta $p_\mu$
conjugate to $x^\mu$. We  will  interpret functions of  $(x,p)$ which are smooth in $x$ and polynomial in $p$ as symbols 
of differential operators acting on ``wave functions''  of $x$. 
 The  $*$-product  will denote
 the operator composition in terms of (Weyl) symbols\foot{Here $ \hbar$ is a formal parameter that  can be always set to 1.
 Note also that we  shall use $\mu,\nu,...$ for coordinate indices and $a,b, ...$ for tangent space indices.} 
\begin{equation}\la{20}
 *=\exp\Big[\frac{\hbar}{2}(\dr{x^\mu}\dl{p_\mu}-\dr{p_\mu}{\dl{x^\mu}})\Big]\,.
\end{equation} 
Let us consider  a   generic  relativistic particle  Hamiltonian 
generalizing the free one $H_0$ 
\be \la{21} 
H(x,p)= H_0(x,p)  +   h(x,p) \ , \qquad \qquad   h(x,p)= 
\sum_{s=0}^\infty  h^{\mu_1\ldots\mu_s}(x)\  p_{\m_1} ... p_{\m_s} \ee
 and subject it to the following
gauge transformations~\cite{Segal:2001di,Segal:2002gd}\foot{Here the 
commutator $[\ ,\ ]_*$   and anticommutator $\{\ ,\ \}_*$  are defined  with respect to the above $*$-product.}
\begin{equation}
\label{gtr}
 \delta H=\hbar^{-1}\qcommut{H}{\epsilon(x,p)}+\scommut{H}{\omega(x,p)}_* \ , 
\end{equation} 
where $\epsilon,\omega$ are unconstrained symbols interpreted as gauge parameters.\foot{This  gauge symmetry is reducible:
$ \delta \epsilon=\sqcommut{H}{\alpha}\,, \quad \delta \omega=-\qcommut{H}{\alpha}$
where $\alpha$ is a reducibility parameter.}
They 
induce the linearized  CHS gauge transformations of the  coefficient  fields $h_s$. 

These gauge symmetries have a simple interpretation~\cite{Grigoriev:2006tt,Bekaert:2013zya} in the   context  of a constrained system
 $T_a(x,p)=0$  where the symbols  $T_a(x,p)$ are 
  subject to the 1-st class condition \\
$ \qcommut{T_a}{T_b}=U_{ab}^c *T_c\,.$
A  given constrained system  can be described by an equivalent set of constraints: an infinitesimal equivalence relation
$T_a\sim T_a +\hbar^{-1}\qcommut{T_a}{\chi}$
corresponds  to  an  infinitesimal canonical transformation while the equivalence relation
$T_a\sim T_a + \lambda_a^b*T_b$
corresponds to an  infinitesimal redefinition of the constraints (which preserves the constraint surface).
Then the space of gauge-inequivalent configurations   is a moduli space of
constrained systems that have fixed number of 1-st class constraints and satisfy certain extra conditions (e.g.   belong to a vicinity of certain vacuum $H_0$).
\footnote{In the   BRST description the constraints are encoded in the symbol $\brst(x,p,\text{ghosts})$ while 
the above equations and gauge transformations are encoded in
$ \qcommut{\brst}{\brst}=0\,, \ \  \delta \brst=\hbar^{-1}\qcommut{\brst}{\Xi}$
where $\Xi$  contains  $\chi$ and $\lambda^a_b$ and $\qcommut{\ }{\ }$ is a graded commutator. In the context of
string field theory  such  sort of systems were considered in~\cite{Horowitz:1986dta}.}

To relate  this   to \rf{gtr}  let us   consider  the case of just one constraint $T\equiv H$ 
 and identify parameters as
$\epsilon=\chi-\frac{\hbar}{2}\lambda$,\ \  $\omega=\half\lambda$. 
Then the gauge transformations \rf{gtr}   are the natural equivalence transformations of the constrained system describing a  
relativistic particle. The ``vacuum"   (quadratic in $p$)  choice of $H$ 
 \be \la{7}
 H_0=g(x,p) \equiv-\half g^{\mu\nu}(x)\  p_\mu p_\nu\ee
  is the standard Hamiltonian 
  of a  particle in a gravitational  background.
In this case the linearized gauge transformations \rf{gtr}  read as 
\begin{multline}\la{8}
 \delta h =\hbar^{-1}\qcommut{H_0}{\epsilon(x,p)}+\scommut{H_0}{\omega(x,p)}_*
 \\ =
 p_\mu g^{\mu\nu}\dl{x^\nu}\epsilon-p_\mu(\half \d_\rho g^{\mu\nu})p_\nu \dl{p_\rho}\epsilon
 -\half g^{\mu\nu} p_\mu p_\nu \ \omega+O(\hbar)\,.
%
%
\end{multline} 
Let $\hat H(x,\dl{x})$  be a differential operator associated to the 
 symbol $H(x,p)$ (assumed  to be such that $\hat H$ is formally hermitian).
As we are using the Weyl symbols
 this means that $H$ is real if $x^\mu$  is real and $p_\mu$ is  imaginary. Then the  complex scalar  action
defined as\foot{
Here ${}^*$ is  complex   conjugation   which should not be confused   with star  product   defined above.}
\begin{equation}
\label{210}
 \SS[\p,h]
 =\int d^d x \ \phi^*(x)\  \hat H(x,\dl{x})\ \phi(x)
\end{equation} 
is invariant under the transformations~\rf{gtr} provided at the same time $\phi$ transforms as
\foot{Note that by removing (anti)hermiticity conditions on $\hat H, \hat \omega,\hat \epsilon$ one finds 
 the ``equations of motion"  version of the system. Indeed, in this case
the above gauge symmetries are symmetries of the following equations of motion:
$ \hat H \phi=0$. 
}
\begin{equation}
\label{211}
\delta\phi=-(\hbar^{-1}\hat\epsilon+\hat\omega)\phi\,, \qquad 
\epsilon^\dagger=-\epsilon, \quad \omega^\dagger=\omega\,.
\end{equation} 
As follows from~\rf{gtr} and properties of the Weyl star product,  one can consistently put to zero all fields $h_s$ 
 of odd spins appearing  in \rf{21} 
  along with the  gauge parameters $\epsilon$ / $\omega$  of even / odd  degree in $p_\mu$.
On top of this there is a consistent truncation to a system  where all fields with $s>2$ and their associated gauge parameters are set to zero.
 This is due to  the fact that the elements which 
 {are} 
 at most linear in $p_\mu$ form a Lie subalgebra of a Weyl star-product algebra. The above two truncations can be combined,
  resulting in a system for fields of spins $0$ and $2$  only. Furthermore, the  spin $0$  field 
 can  also  be  eliminated. 
Apart from the above consistent truncation to spins $\leq 2$ 
one  can not  get a gauge invariant action  depending only on  a finite number of fields $h_s$. 

\subsection{Conserved currents}

For   $H$ in \rf{21},\rf{7}  
 the action \rf{210} may be written as  
\begin{equation}\la{23}
 \SS=\int d^d x \left[\phi^*\, \hat H_0\,  \phi +  \phi^*\,  \hat h\,  \phi \right]\,.
\end{equation} 
The condition of its   invariance under~\rf{gtr}  combined with $\delta \phi=-\hbar^{-1}\hat\epsilon\, \phi$ takes the form
\begin{equation}
\label{24}
 \int d^dx \left(\big(\commut{H_0}{\epsilon}_*+
 \commut{h}{\epsilon}_* \big)\vddl{\SS}{h}-\hat\epsilon\,  \phi^* \vddl{\SS}{\phi}\right)=0  \ . 
\end{equation} 
Introducing a   generating function for conserved currents 
(here $u^\m$  is an auxiliary constant  vector)
\be \la{25} 
J=\sum_s  \frac{1}{s!}u^{\mu_1}\ldots u^{\mu_s} \vddl{\SS}{h^{\mu_1\ldots\mu_s}}\ , 
\ee
the quadratic in the  fields  term in  \rf{24}  may be written as 
\begin{equation}\la{155} 
\int d^dx \left(\inner{\qcommut{H_0}{\epsilon}}{J}   - 2({\hat\epsilon\phi})^* { \hat H_0 \phi}\right)=0\,,
\end{equation} 
where $\inner{\ }{}$ denotes a  natural inner product (contraction of indices) between polynomials in $p_\mu$ and polynomials in $u^\mu$.\footnote{For instance,  $\inner{1}{1}=1$, $\inner{u^\mu}{p_\nu}=\delta^\mu_\nu$, etc. Note that then $(u^\mu)^\dagger=\dl{p_\mu}$,  etc.}
For $H_0=-\half \eta^{\mu\nu} p_\mu p_\nu\equiv  - \ha  p^2 $ one gets the usual on-shell conservation condition for the   currents 
\be\la{15} \eta^{\mu\nu}\dl{u^\mu}\dl{x^\nu} J=0  \ . \ee
Applying analogous arguments  to the second gauge invariance with the parameter $\omega$ in \rf{gtr} 
results in the  generalized on-shell tracelessness condition for the currents:
\begin{equation}
 \int d^dx \left(\inner{ { \scommut{H_0}{\omega}}_*}{J} - 2({\hat\omega\phi})^*{\hat H_0 \phi}\right)=0\,.
\end{equation} 
For $H_0=-\half p^2$ one gets the ``deformed" tracelessness condition
\begin{equation}
\label{deftr}
 (\eta^{\mu\nu}\dl{u^\mu}\dl{u^\nu}-\half \hbar^2 \Box)J=0\,.
\end{equation} 
Redefining the components of $J$ one can make them strictly traceless ~\cite{Segal:2002gd}   but 
 it is not always   useful to  perform this  redefinition explicitly.

More generally, taking a variational derivative of \rf{24} with respect to $\epsilon$ and not decomposing 
the result according to  the homogeneity in the fields  leads to a 
nonlinear generalization of the conservation condition \rf{15} 
which now involves  the fields $h_s$. Analogous arguments apply to
 gauge invariance under the transformations with the parameter $\omega$
  leading  to a nonlinear version  of~\rf{deftr}.


\section{Covariant   form of  
 the   scalar  field  action in CHS background }
\label{covariant-q}

To generalize the above discussion  to a curved background  we shall first   consider a   covariant framework 
for a relativistic particle quantization. 
One can naturally define a quantization of the cotangent bundle over a  curved  spacetime in a geometrically covariant way. This  covariant description is based on 
 the metric  $g_{\mu\nu}$   and the  metric connection  and 
 includes (see also Appendix~\bref{sec:fedosov}):
\begin{itemize}
 \item {\it star product}: for any functions of $x^\mu,p_\nu$ which are smooth in $x$ and polynomial in $p$ there is a well-defined 
 (and unique
  under some extra  natural conditions)
   associative $*$-product. 
 The $*$-commutator (anti-commutator) carries odd (resp. even) homogeneity in $p$.\foot{For example,   if $f(x,-p)=f(x,p)$ and $u(x,-p)=u(x,p)$ then $\qcommut{f}{u}(x,-p)=-\qcommut{f}{u}(x,p)$.}
 \item {\it state space}: space of functions of  $x^\mu$  equipped with the natural inner product
\begin{equation}\la{31}
 \inner{\phi}{\chi}=\int d^dx \sqrt g \,\, \phi^*(x)\chi(x) \ . 
\end{equation} 
 \item {\it symbol map}: there is a well-defined map $f \to \widehat f$ 
  from functions of $x,p$ (symbols) to differential operators acting on   
 functions of  $x^\mu$  
 (``wave functions") such that 
 \begin{equation}\la{323}
  \widehat{f_1* f_2}=\hat f_1\,\, \hat f_2 \,.
 \end{equation} 
The operator associated to  $f(x,p)$ is denoted by $\Rho(g,f)=\hat f(x,\dl{x})$. Here  $g$ indicates the dependence on the background metric,  i.e. on 
{the} covariant derivative and {the} curvature  built out of it  (see Appendix~\bref{sec:fedosov} and more specifically~\rf{symb-map} and propositions~\bref{f-lift} and \bref{phi-lift}). The real symbols correspond  to 
hermitian operators.  
 \end{itemize}
 

Let us redefine the   spin 0 part of $h(x,p)$  in \rf{21}  by the  scalar curvature of the metric and write $H$ in \rf{21},\rf{7}  as 
\be
\label{32}
H(x,p)=g(x,p) +{\cR}(x) + h(x,p)\,, \qquad g(x,p)= -\half g^{\mu\nu}(x) p_{\mu} p_{\nu}\,, \qquad \cR=\gamma R\,,
\ee
where $R$ is  the scalar curvature   and $\gamma$ is a  numerical coefficient. 
The covariant version of the scalar field action \rf{210},\rf{23} 
then  reads 
\begin{equation}
\label{3.4}
 \SS[g,h,\phi]=\int  d^d x \, \sqrt g \,\, \phi^*\left(\widehat{g(x,p)} + \cR(x)\right)\phi+
\int  d^d x \, \sqrt g \,\, \phi^* \hat{h(x,p)}\phi\,,
\end{equation} 
where we have explicitly separated the $h$-independent term.
%
The coefficient $\gamma$ in \rf{32} is chosen  so that the first term in the above action is the standard  action of the conformally coupled scalar.
  Note that this  action  depends on  $g_{\mu\nu}$ also through the metric connection entering the symbol map.

By construction,  this action is invariant under the covariant version of~\rf{gtr}  and \rf{211}\foot{Below 
 we set $\hbar=1$ for notational   simplicity.}
\begin{align}
\label{355}
& \delta h=
\qcommut{g+\cR+h}{\epsilon}+\scommut{g+\cR+h}{\omega}_*\,, \\ 
& \delta \phi=-\widehat{( 
\epsilon+\omega)}\phi\,.  \la{3555}
\end{align} 
Let us stress that the background field $g_{\mu\nu}(x) $  is not affected by this gauge transformation. However, there are hidden gauge transformations of the action \rf{3.4}
 related to redefinition of $g^{\mu\nu}$ and 
$h_2 = (h^{\mu\nu})$ which do not change their sum $g+h$ modulo
relevant redefinition of the symbol map (which depends on $g^{\mu\nu}$). For 
 further analysis it is useful to employ  two  extra types of symmetries which have natural geometrical meaning
and are,  in fact,  certain combinations of \rf{355},\rf{3555} and  redefinitions of $g^{\mu\nu}$ and $h^{\mu\nu}$.

First, 
given that  the  action  \rf{3.4} contains only  covariant objects, it is invariant under the diffeomorphisms generated by a vector field $\xi=\xi^\mu \dl{x^\mu}$   with $h^{\m_1...\m_s}$ transforming as tensors, i.e. under 
\begin{equation}\la{35}
 \delta g=
  {\cal L}_\xi g\,, \qquad\qquad  \delta h ={\cal L}
 _\xi h \,, \qquad\qquad  \delta \phi=\xi\phi\ .
\end{equation} 
Second, the $h$-independent 
 term in~\eqref{3.4} is the action of a conformally coupled scalar and hence it is invariant under the usual Weyl symmetry $\delta_{\omega_0} g^{\mu\nu}=2\omega_0 g^{\mu\nu}\,,\quad  \delta_{\omega_0} \phi= (\frac{d}{2}-1) \omega_0 \phi$, where $\omega_0$ is $p$-independent. The second term in \rf{3.4} 
 can also be made invariant by setting $\delta_{\omega_0} h =2\omega_0 h+ \delta^{\prime}_{\omega_0} h$, where 
\begin{equation}
\label{delta-prime}
\delta^{\prime}_{\omega_0} h = \scommut{\omega_0}{h}_* -2\omega_0 h+\delta^{\prime\prime}_{\omega_0} h\,.                                                                                                      \end{equation}  
Here $\delta^{\prime\prime}_{\omega_0} h$ takes into account the variation of the symbol map under the variation of the metric.  More precisely, because the map between the operators and the symbols is one-to-one, one can always trade a variation of the symbol map for an appropriate variation $\delta^{\prime\prime}_{\omega_0} h$ of $h$. 
  If we denote by $\Rho(g,h)$ the operator associated by the symbol map to the  symbol $h$ then 
\begin{equation}\la{3.8} 
 \Rho(g+2\omega_0 g, h)=\Rho(g, h+\delta^{\prime\prime}_{\omega_0} h)+O(\omega_0^2) \ . 
\end{equation} 
One can then  represent  the variation of $h$ as
$\delta_{\omega_0} h =\scommut{\omega_0}{h}_*+\delta^{\prime\prime}_{\omega_0} h= 2\omega_0 h+\delta^\prime_{\omega_0} h$.
It follows from the structure of the star product and the symbol map that $\delta^\prime_{\omega_0}   h_s $ is linear in  the CHS fields 
and only depends on $h_{s_i}$ with $s_i>s$.\footnote{
The only nontrivial pont is to check this for $\delta^{\prime\prime} h_s$. 
The terms in the $\hat h \phi$ involving $s$ derivatives of $\phi$ have the structure
$h^{\nu_1\ldots \nu_t}\d_{\mu_1}\ldots \d_{\mu_s} \phi$ where  contractions of indices between  the two groups  may be via  $\delta$-symbol, $\Gamma^\mu_{\nu\rho}$, and the curvature and its derivatives.
It is clear that any such contraction can be nonvanishing only for $t\geq s$
(note the number of  upper and lower indices in $\Gamma,\, R$ etc.), and,  moreover, 
at $t=s$ one  can have  the leading contribution   where only $\delta$-symbols are employed in the contraction. Furthermore, the variations of the above expression under the change of
$g^{\mu\nu}$ and the respective change of the connection, curvature, etc.,  can be compensated
by the variation of $h_s$. This way one finds the compensating transformation  $\delta^{\prime\prime}_{\omega_0} h_s$ which,  by construction,  is  proportional to $h_t$ with $t>s$.
For instance, $\delta^\prime h^a$ contains the terms such as $h^{abc}\nabla_b \nabla_c \omega_0$, $\delta_{\omega_0}(\nabla_c h^{ca})$,
$\delta_{\omega_0}(\nabla_b\nabla_c h^{bca})$ as well as further terms involving $h_2,h_3,\ldots$. Here, $\delta_{\omega_0}$ 
acts only on the Levi-Civita connection coefficients and the respective curvature and denotes their variation 
under the Weyl transformation.}

We conclude that the action \rf{3.4}  has  an  infinitesimal symmetry  which is the direct analog of the usual Weyl  transformations 
\begin{equation}
\label{3.7}\te 
 \delta_{\omega_0} g^{\mu\nu}=2\omega_0 g^{\mu\nu}\,, \qquad \qquad \delta_{\omega_0} h =2\omega_0 h+ \delta^{\prime}_{\omega_0} h 
  \,, \qquad \qquad 
 \delta_{\omega_0} \phi= (\frac{d}{2}-1)\omega_0 \phi \,.
\end{equation}
It will be called  the deformed Weyl symmetry in what follows.


\section{Covariant  expansion of the CHS action in a non-trivial metric }

Starting with  the covariant version~\rf{3.4} of the scalar field action minimally coupled to the CHS  fields 
one can integrate out the scalar $\phi$ and extract the  local log-divergent part $S[g,h]$ of the resulting induced  action  as in \rf{133}.
 This local term  is  invariant  under the $h$-field  part \rf{355}  of the gauge symmetries  \rf{355},\rf{3555}
 {as well as under the symmetries  \rf{35},\rf{3.7}}  of the original scalar action 
  and thus provides a natural definition  of the  CHS action $S[g,h]$  in a general metric background.

\subsection{Expansion of  the CHS action}


Let us specify to the case of $d=4$  and  consider the expansion of $S[g,h]$ in  powers of  $h_s$
\begin{align}
\label{310}
 & S[g,h]=S[g]+S_1[g,h]+S_2[g,h]+\ldots\,,\qquad \\
 & S_1[g,h]=\sum_s
  \int d^4 x \sqrt{g} \,K_{\mu_1\ldots\mu_s}[g]\ h^{\mu_1\ldots\mu_s}\,, \qquad  
  S_2 =\sum_{s,s'}  \int \  h_s \ O _{ss'} [g] \ h_{s'} 
  \ , \ \ ...
  \la{3100}
 \end{align}
Here we ignore total derivatives and hence $K_s=(K_{\mu_1\ldots\mu_s})$ can be assumed to be 
a local function of the metric $g$. 
The diffeomorphisms~\rf{35} transform $g $ and $h_s$ through themselves. Under the  deformed Weyl transformations \rf{3.7} 
$g$  gets rescaled    while  $h_s$  transforms into  $h_t$ with $t\geq s$.
As   $S[g]$ must be invariant under both diffeomorphisms
and the usual Weyl transformations of the metric $g$  and is local,  it should be 
 the standard $C^2_{\m\n\lambda\rho}$ Weyl action (cf. the discussion of  Weyl invariants  in Appendix~\bref{sec:weyl}).

As the diffeomorphism and the deformed Weyl symmetries are homogeneous in $h$, the linear in $h$ term  $S_1$ 
 must be invariant on its own. 
 Thus $K_{\mu_1\ldots\mu_s}[g]$ should be 
  a tensor under the diffeomorphisms and should vanish (or give a total derivative) if $g_{\mu\nu}$ is flat.
The  fact that the  flat-space CHS action has no terms linear in $h$  is  clear  directly from \rf{13},\rf{133}. 
Indeed, as $h_s$ has  mass dimension $2-s$   and the CHS action is  local and dimensionless,
 $K_s$  in $S_1$  should have dimension $2+s$, i.e. it should   have a structure
   $\nabla^{2+s} +  R \nabla^s  + .... + R^{ 2+s \ov 2}$  where $R$  is the curvature.
We   shall ignore  the   leading  highest derivative  term  as   it  gives a total derivative   in \rf{3100}.

Let us note that  for a  flat $g_{\m\n} $  background  the   quadratic in $h_s$  term is  not manifestly diagonal  before one performs  the  algebraic redefinition of the fields  (that takes 
 care of the traces of the fields, i.e.  is related to  the algebraic part of the   gauge transformations \ci{Segal:2002gd}).
   For a non-trivial $g_{\m\n} $   one will face a more serious  non-diagonality issue  due to terms  involving the curvature  of $g_{\m\n} $  that 
   mix  fields of different spin; this is   related  to the   differential part of the gauge transformations.

Suppose that $g^{\mu\nu}=g_0^{\mu\nu}(x)$ and  $h_s=0$  (for all  $s$)  is a particular solution of the 
equations  corresponding to  $S[g,h]$.  
 The   necessary and sufficient conditions for that 
are  (ignoring total derivative terms)  
 \begin{equation}
 \label{k0}
  \vddl{S[g]}{g}\Big|_{g=g_0}=0
  \,,\qquad \qquad K_{\mu_1\ldots\mu_s}[g_0]=0\,.
 \end{equation} 
 Thus   $g_0$  should   be Bach-flat   and $K_s$ should   vanish on a Bach-flat  background. 
Then the  expansion of \rf{310}   near this solution 
 reads as
\begin{equation}
 S[g_0, h]= S[g_0] + S_2[g_0,h] + ...\,, \la{313}
\end{equation} 
where we set to  zero  the perturbation $\bar h^{\m\n}$ of $g^{\mu\nu}$ itself. As 
$g^{\mu\nu}$ is not affected by the gauge transformations  \rf{355},\rf{3555} 
 the term  $S_2[g_0,h]$
should   be  invariant under the linearized  version of \rf{355}, i.e.
\begin{equation}
\label{lings}
\begin{gathered}
\delta h_s=\Big(\qcommut{g_0+\cR_0}{\epsilon}+\sqcommut{g_0+\cR_0}{\omega}\Big)\Big|_s\ . \quad 
\end{gathered}
\end{equation}
Here $A\big|_s$ denotes the projection to spin $s$ of the generating function $A(x,p)$, i.e. the term of homogeneity $s$ in $p_\mu$.

It follows from the structure of the star-product that one can consistently put to zero all the fields with $s >s_0$ along with $\epsilon$ parameters of homogeneity $>s_0-1$ and $\omega$ of homogeneity $>s_0-2$ in $p$. 
Hence $S_{2}[g_0,h]\big|_{h_{s>s_0}=0}$
is invariant under  the linearized gauge transformations~\rf{lings} with $\epsilon$ of degree $<s_0$ and $\omega$ of degree $<s_0-1$.



Let us now show the vanishing of $K_s[g_0]$ in Bach-flat   background 
  at least to first order in the  background  curvature. 
This will  generalize  the result of~\cite{Nutma:2014pua}  that the $s=3$  kinetic   operator  is gauge-invariant  in a Bach-flat background to linear order in  the  curvature. 

\subsection{Conditions for the vanishing of the linear  fluctuation  term }
\label{sec:K-vanish-lin-order}

Let us study  the consequences of the deformed  Weyl  symmetry \rf{3.7}
 for  the structure of $K_{\mu_1\ldots\mu_s}[g]$.
 It  is useful to introduce the following notation
for the  transformation of $K_s$ under $\delta g^{\mu\nu}=2\omega_0g^{\mu\nu}$:
\begin{equation}\la{317}
\delta^\prime_{\omega_0} K_{\mu_1\ldots \mu_s}=\delta_{\omega_0}K_{\mu_1\ldots \mu_s}- 2\omega_0 K_{\mu_1\ldots \mu_s}\ . 
\end{equation} 
Consider the variation of $\sqrt{g} K_s h_s$ under the deformed Weyl transformation.  Since $\delta^\prime h_0$ 
(defined in~\eqref{delta-prime})
 does not  depend on $h_0$, the only term proportional to $h_0$ is $\delta^\prime K_0 h_0$ and hence $\delta^\prime K_0=0$. 
This implies that $K_0$ is Weyl invariant with weight $-2$ (i.e. behaves like $g^{\mu\nu}$)
but there are no such non-trivial  invariants (see ~Appendix~\bref{sec:weyl}).


Let us proceed by induction. Assuming that  we have shown that  $K_r=0$ for $r<s$, let  us  consider the variation
of $\sqrt{g} K_s h_s$ under the deformed Weyl transformation. Concentrating on the terms in the variation proportional to $h_s$ gives
\begin{equation}
 (\delta^\prime_{\omega_0}K_s )h_s+ \sum_{l=1}^\infty K_l (\delta^\prime h_l)\big|_{h_t=0, t\neq s}=0\,.
\end{equation} 
Note that $(\delta^\prime h_l)\big|_s=0$
for $l\geq s$.  Taking this and $K_l=0$ for $l<s$ into account one finds $\delta^\prime_{\omega_0}K_s=0$ and hence $K_s$
is a tensor which is  Weyl invariant of weight $-2$. 

Next, let us  consider the gauge transformations~\rf{355}. The gauge variation of $S_1$ under the inhomogeneous in $h$ part $(\qcommut{g+\cR}{\epsilon}+\sqcommut{g+\cR}{\omega})\big|_s$ of~\rf{355} should vanish. Setting  to zero all the fields of spin $>s$ and the associated gauge parameters one gets
\begin{equation}
\label{covcons}
 \int d^4 x \sqrt{g}\,  K_{\mu_1\ldots \mu_s} \nabla^{(\mu_1}\epsilon^{\mu_2\ldots \mu_s)}=0\,, \qquad\qquad  \int d^4x  \sqrt{g}\,  K_{\mu_1\ldots \mu_s} g^{(\mu_1\mu_2}\omega^{\mu_3\ldots \mu_s)}=0\ .
\end{equation} 
This  leads to 
\begin{equation}\la{4.9}
 \Tr\,  K_s=0 \ ,\qquad  \qquad \nabla^{\mu_1} K_{\mu_1\mu_2\ldots \mu_s}=0\ ,	
\end{equation} 
i.e. $K_s$  should have  the  same properties   as a  covariantly-conserved traceless current. 
The relations \rf{covcons}  and \rf{4.9}   have direct generalizations to $d >4$  dimensions. 

We have thus shown that under the induction assumption $K_s$   should be  a Weyl invariant   tensor
of weight $2$ which is also traceless and covariantly conserved. The totally symmetric
traceless tensors of Weyl weigh $-2$ are  called  in the math literature
as ``admissible invariants". For $s\leq 3$ these invariants are known explicitly~\cite{LeBrun:1990}: for $s=1$ any invariant vanishes;
for $s=2$ it is proportional to the Bach tensor; for $s=3$ it is proportional to the Eastwood--Dighton tensor 
(see Appendix~\bref{sec:weyl} for more details):
\begin{equation}\label{ed}
E_{\mu\nu\rho}\to E_{ABCA'B'C'}=\Psi_{ABCD} \nabla^{DD'}\Psi_{A'B'C'D'}-
\,\Psi_{A'B'C'D'}\nabla^{DD'}\Psi_{ABCD}\,.
\end{equation} 
Here we resorted to the spinor conventions  where $\Psi_{ABCD}$ and $\Psi_{A'B'C'D'}$ are Weyl spinors
corresponding to (anti)self-dual components of the Weyl tensor. 

{For  general $s$,  one may consider a special case of 
4d  Bach-flat backgrounds  with  Weyl tensor  which  is (anti)selfdual. 
It is easy to see that in this case $K_s$ must vanish (see~Appendix~\bref{sec:weyl}) and hence CHS are consistent.}

{Considering  generic  backgrounds,} 
let us  restrict attention to terms  in $K_s$    which are linear in the curvature or Weyl tensor $C$. Then  
we will have $\nabla^s C$ like terms where  $s$ indices are symmetrised and $4$ indices are contracted by $g^{\mu\nu}$. Since  $\nabla^s C$  should be  a  totally symmetric tensor and since $C$ has 4 indices and is traceless, two of the derivatives   should act on $C$ itself.  Such terms should vanish on a Bach-flat  background.

It would be  important    to extend the  above argument of the vanishing of $K_s$ 
 beyond the linear  in 
curvature terms. Let us make few comments  that may be useful   for an attempt to prove this. 
  As the  metric  $g$    should actually be a  background   for the spin $2$  field $h_2$, there   should be  a hidden gauge symmetry which transforms $g$ and $h_2$
  in such a way that their sum $g+h_2$ remains invariant while  all other   fields   $h_s$ 
   also transform to compensate for the change of the symbol map.
    This symmetry  may be useful to eliminate some unwanted terms. Another remark  is that 
  we are dealing with the CHS theory involving an infinite set of  fields  but     so far 
   made use of  only some  of the gauge symmetries that preserve the subspace of field configurations where only a finite collection of fields are non-vanishing. In particular, for the above arguments to work it is enough to  compute  the  CHS action as the divergent part of the scalar effective action \rf{133} with  $h_r=0$ for $r>s$. 
   This way one may  avoid subtleties related to the
   fact that the full space of CHS fields  is  infinite-dimensional. 
   Finally, let us mention that the entire construction can probably {be} made more geometrical by employing the conformally equivariant quantization 
    which is known~\cite{Radoux:2009,Silhan:2009} for generic conformal manifolds.


%



\subsection{Gauge invariance of  
spin-$s$    quadratic term
 to first order in curvature}

As we have  argued above,  $K_s[g_0]$  must vanish  at least up to terms of second order  in the curvature\foot{By curvature terms 
 we always  mean the products of  Riemann tensor and its covariant derivatives.} 
  if 
the background metric  $g_0$ is Bach-flat. The gauge invariance of the complete action~\rf{310} at the zeroth and the first order in $h$ gives
\begin{equation}
\label{01-exp}
 \int \sqrt{g}\,K_s\delta^0_\epsilon h_s=0\,, \qquad \int\sqrt{g}\,   K_s\delta^1_\epsilon h_s+\int\sqrt{g}\, \vddl{S_2}{h_s}\delta_\epsilon^0 h_s=0\,,
\end{equation} 
where $\delta^0$ and $\delta^1$ denote the leading 
 and the linear in $h$ parts of the gauge transformation. 
 As  $K_s\sim  C^2$ 
(here $C$ denotes the Weyl tensor and its Weyl-covariant derivatives, see Appendix~\bref{sec:weyl}) the second equality implies 
\begin{equation}
\label{1st-ord-inv}
 \delta^0_{\epsilon,\omega} S_2[g_0,h]=O(R^2)\,,
\end{equation} 
where  $\delta^0_{\epsilon,\omega}$ denotes the gauge transformation linearized around $g=g_0,\ h_s=0$.
To zeroth order in the curvatures the gauge transformations are  explicitly
\begin{equation}
\label{0th-curv-gs}
 \delta^0_\epsilon h_s=(p_\mu\nabla^\mu) \epsilon_{s-1}, \qquad \te
 \delta^0_\omega h_s=-\half g^{\mu\nu}p_{\mu}p_{\nu}\, \omega_{s-2}-\frac{1}{4}g^{\mu\nu}\nabla_\mu\nabla_\nu \omega_{s} \,.
\end{equation} 

{We now set to zero all the fields with $s>s_0$ and their associated gauge parameters. For $s=s_0$
\eqref{0th-curv-gs} then gives the exact linearized gauge transformation. Indeed, the curvature contributions may only affect fields
 with the spins lower than $s_0$. Note also that the second term in the expression for $\delta_\omega h_s$ vanishes 
  for $s=s_0$.
 Moreover, to zeroth order in the curvature this term can be removed~\cite{Segal:2002gd} by the field and the gauge parameter redefinition.}
Upon this redefinition the gauge transformation takes the standard diagonal form, with the usual derivatives replaced by the covariant ones. This implies that to zeroth order in the curvature the term 
$S_2[g_0,h]$ is just a direct sum of the standard quadratic actions for all spins $1,2,\ldots,s_0$.

Let us now  include terms of first  order in the curvature. Because to zeroth order in curvature
$S_2[g_0,h]$ is diagonal (does not contain terms mixing different spins) the gauge invariance implies
that the quadratic in $h_{s_0}$ term in $S_2[g_0,h]$  is gauge invariant on its own up to terms of second order in curvature. 
{Indeed, under the transformation with only $\epsilon_{s_0-1}$ and $\omega_{s_0-2}$ nonvanishing
the terms in the variation of $S_2[g_0,h]$ that are linear in $h_{s_0}$ either originate from the quadratic in $h_{s_0}$
term or from the variation of $h_s$ with $s<s_0$ in the mixing terms. In the later case the variation is at least of order $2$
so that~\eqref{1st-ord-inv} implies the assertion.} This generalizes the spin $3$ statement from~\cite{Nutma:2014pua} to  any   integer spin case. 

Let us note that, strictly speaking,  in the above considerations we made use of the expansion in Riemann curvature  while the 
vanishing of $K_s$ was shown to first order in the Weyl curvature. This is the same  for special case of Ricci-flat 
backgrounds, but there are Bach-flat backgrounds that are not Ricci-flat. 
Taking into account the (deformed) Weyl invariance it should be possible to demonstrate  also 
the gauge invariance to first order in  the Weyl curvature.

\subsection{
Spin 3 example }

Let us now assume that the background metric is chosen such 
 that  both the  Bach tensor \rf{55}   and the Eastwood-Dighton tensor \rf{ed}  vanish, i.e.
  $B_{\mu\nu}=0=E_{\mu\nu\rho}$. For an algebraically-general Weyl tensor this
implies that the metric is conformally Einstein~\cite{Kozameh1985,LeBrun:1990}.
Unfortunately, the vanishing of $K_3$ in \rf{3100},\rf{4.9}  does  not  directly  imply 
that the spin $3$  CHS field kinetic term  is   always consistent (i.e. gauge-invariant)  on such a background. 
Taking $\epsilon=\epsilon^{ab}p_ap_b$
and extracting the linear in $h_3$ contribution in
 the $h_4$ variation in the  second equation in~\eqref{01-exp} one gets:
\begin{equation}
\label{K4s3}
\int d^4x \sqrt{g}\, K_4 \qcommut{h_3}{\epsilon_2}\Big|_4+ \int d^4x\sqrt{g}\, h_3\frac{\delta^2 S_2}{\delta h_3 \delta h_1}\delta^0_{\epsilon_2} h_1+\int  d^4x\sqrt{g}\, h_3 \frac{\delta^2 S_2}{\delta h_3 \delta h_3} \delta^0_{\epsilon_2} h_3=0
\ . \end{equation} 
Thus 
 if $K_4$ were  nonvanishing beyond the leading order in curvature,  our argument   would  not in general 
 imply that the spin 1 plus  spin 3 system is consistent on its own.
 Below we  shall  assume  that this is not the case, i.e. all $K_s$  vanish on  Bach-flat background to all orders in curvature expansion.

It is clear from the structure of~\eqref{K4s3} that on a non-trivial   background 
the spin 3 field   may mix  with the spin 1 in the  quadratic term $S_2$ in \rf{310}. 
To understand the reason  for this mixing 
let us  go back to the discussion in sections 2 and 3    and 
consider the linearized gauge transformations 
around the  vacuum  Hamiltonian 
$H_0= -\half g^{\mu\nu}p_\mu p_\nu=-\half \eta^{ab}p_a p_b$ in \rf{7}.
As was noted above,  it 
 follows from the structure of the star-product that the linearized gauge transfomations with parameters $\epsilon$ and $\omega$ of degree $s-1$ and $s-2$ respectively can only affect the  fields of spins $s,s-2,s-4, \ldots$. 
 Thus the simplest nontrivial system is that of
spins $1$ and $3$. 
For $s=1$  field the gauge transformations are standard.  For  $s=3$ the  parameters are 
 $\epsilon^{ab}p_a p_b$ and $w^a p_a$. Let us first consider the  gradient-like transformation 
\begin{equation}
\label{gs-3-3}
 \delta (h^{abc}p_a p_b p_c)=\commut{-\half \eta^{ab}p_a p_b}{\epsilon^{cd}p_cp_d}_*{}\Big|_{3}=p_a p_b p_c (\nabla^a\epsilon^{bc})\ , 
\end{equation} 
where we projected to the spin 3 component and {disregarded the contribution from the background scalar curvature.} This is thus  a natural covariantisation of the  flat-space  gradient gauge transformation. However, the  transformation generated by $\epsilon^{cd}p_c p_d$ gives  also a  non-zero contribution to the spin $1$ sector:\foot{Let us stress that {these} relations  are complete  as terms of degree higher than 2 in the covariantly constant lifts of $\eta$ and $\epsilon$ can not contribute.
For comparison, in the sector   of spin 2 and  spin 0 fields   with 
$H=\eta^{ab}p_a p_b+h_0(x)$,  the gauge transformations with parameters $\epsilon=\epsilon^a(x) p_a$
and $\omega=\omega(x)$ read  (restricted to  $h_0$, cf. also ~\cite{Segal:2002gd})
\begin{equation}\no\te 
 \delta h_0=\epsilon^a \nabla_a h_0 + 2 \omega h_0+\frac{
 1}{2}\eta^{ab}\nabla_a\nabla_b \omega \,.
\end{equation} 
Using the transformation law of the scalar curvature $\delta_\omega R=2\omega R +\eta^{ab}\nabla_a\nabla_b\omega$ under Weyl transformations of the metric $g \to \omega g$  one finds that $h_0^\prime=h_0- \gamma 
R$ transforms homogeneously:
$
\delta h_0^\prime=\epsilon^a \nabla_a h_0^\prime + 2 \omega h_0^\prime.
$ 
It follows from the  above transformation law that one can consistently put  $h_0^\prime$ to zero in the  scalar  field action
$\int d^dx \sqrt g \big(\phi^* (\hat g+ \hat h_0) \phi\big)=
\int d^dx \sqrt g \phi^* \big(-\nabla^2 + \gamma  
R+h^\prime_0\big)\phi$.
}
\begin{equation}
\label{gs-3-11}
\delta (h^{a}p_a)=\commut{-\half \eta^{ab}p_a p_b}{\epsilon^{cd}p_c p_d}_*{}\Big|_{1}=\te 
-\frac{4}{3}R^a_{bcd}\nabla_a\epsilon^{cd}p^b\ . 
\end{equation} 
Thus the  linearized gauge transformations with parameters $\epsilon^{ab}$ and $\omega^a$  will act on  spin $1$   field as
\begin{equation}
\label{5.4}
\delta h_a= \te 
-\frac{4}{3}R{}_{ab}{}^c{}_d\nabla_c \epsilon^{bd}- \frac{1}{4}\nabla^2 \omega_a
\ . 
\end{equation} 
While the second term here
can be removed by a field redefinition (the ``dressing map" of~\cite{Segal:2002gd})
$h_a\to h_a+ 
c \nabla^2 h_{ab}{}^{b}$,  the first term is non-trivial.\foot{ 
Let us note that  the need for spin 1 field to transform under the spin 3 gauge transformations when the background is not 
conformally flat   can be  concluded  directly  from the study of the conservation condition of a complex scalar spin 3 current on a curved (e.g. 
Ricci-flat)  background (R. Roiban and A. Tseytlin, unpublished). 
   If  $J_s \sim \phi^* \nabla... \nabla \phi + ...$  then  on a   Ricci-flat background  and using $\nabla^2 \phi=0$ one can show that 
$\nabla^\m J_{\m\nu\lambda} \sim C_{\nu \sigma \lambda\rho } \nabla^\sigma J^\rho $
 implying that  to have the $h_\mu J^\mu + h_{\m\nu\lambda}  J_{\m\nu\lambda} $  coupling term to be invariant under spin 3 gauge transformations 
 $\delta h_{\m\nu\lambda} = \nabla_\mu \epsilon_{\nu \lambda} + ...$ one is to modify the spin 1  transformation  by $  C_\rho ^{\ \nu \mu \lambda}   \nabla_\mu  \epsilon_{\nu \lambda} $ term.}

The  presence  of the $ \epsilon^{bd}$ term in \rf{5.4} 
  implies   that  the standard  Maxwell $\partial h_1 \partial   h_1$   term in the quadratic  action  $S_2$ 
can not be invariant 
under such transformation.
As a result,  we should then expect   $h_1      h_3$ mixing, i.e. 
    non-diagonal  terms like 
  $R \nabla h_1 \nabla h_3  +  R  \nabla \nabla h_1\, h_3 +     R R h_1 h_3$   that should   compensate
  for the variation    of  the quadratic  in $h_1$ term 
 under  the $h_3$  gauge transformation  in \rf{5.4}.\foot{Note that  in the constant curvature space where
$ R_{abcd}=\lambda(\eta_{ac}\eta_{bd}-\eta_{bc}\eta_{ad})$
the first term in \rf{5.4}  takes the form
$ \delta h_a=\lambda(\nabla_a \epsilon^b_b-\nabla_b\epsilon^{b}_{a})
\sim  \nabla_a \epsilon^b_b+\eta^{bc}(\nabla_{(a}\epsilon_{bc)}) $
and hence can be removed by a combination of field redefinition and gauge parameter redefinition. 
The same  should  be true  also  for general conformally-flat metrics.} 

\iffa 
XXXXXXXXXX
must be invariant~\footnote{A 1-3 cross-term can only compensate the contribution that can be removed by  a 
 field redefinition. In addition,  we assume that there is no  linear terms in $h_s$.} 
 under the first term in~\rf{1gtr}. 
 Thanks to the invariance under the usual gauge transformation $\delta h_a=\nabla_a \epsilon$ 
  the quadratic in $h_a$ term   should be   the square of the $h_1$ field strength, $F_{ab} F^{ab}$. 
  Under \rf{5.4}   it will transform to $\nabla \nabla h_1  R \nabla \epsilon_2$ term
  which depends on  antisymmetric combination $\nabla_{[b}\epsilon_{c]d}$.
To be able to cancel this  by $\delta  h_{ab} = \nabla_{(a} \epsilon_{bc)} $  variation of the 
mixing term like   $\nabla \nabla h_1  R  h_3$     which  depends on totally symmetric  part of $\nabla_{a} \epsilon_{bc} $
we need to require that 
\begin{equation}\label{418}
 R_{a}^{\ bcd}\nabla_c \epsilon_{db} = \nabla_a \chi+ Q_a^{bcd}\nabla_{(b}\epsilon_{cd)} \ , 
\end{equation} 
i.e. that there exist  some  $\chi[R,\epsilon]$ and $Q_a^{bcd}[R,\nabla]$ such that \rf{418}  holds for any $\epsilon$.
This  may be viewed as a condition on background metric. 
As was   noted  in the previous footnote,    this condition holds for constant curvature spaces
and since $\epsilon_{cd}$   should be traceless, in this case $h_1 h_3$   mixing term need not be  introduced. 
\fi

{As we have  seen above, to first order in
the  curvature the  mixing terms       in the action  like  $h_1 h_3 $  one 
do not affect the gauge invariance of the quadratic  term   $S_2$ 
under the transformation
with parameters $\epsilon_{s-1}$ and $\omega_{s-2}$
(so there is no contradiction with ref. \ci{Nutma:2014pua}   where quadratic in $h_3$ 
  term in the action was constructed  to linear order in the curvature  by imposing the condition of gauge invariance).
However, to second order in the curvature
 the 
mixing terms can not be neglected. Then  
it is natural to expect that in general only a system
of all spins $s,s-2,s-4,\ldots$ can be well-defined on  a  sufficiently
 curved background.}

The presence of non-diagonal   terms  in $S_2$  on  curved background  is thus 
expected in general and 
  deserves further
   study.

   \iffa
   \foot{In particular,  it  is  not    a priori  clear if their presence   may   effect 
    the computation of  conformal anomalies of CHS fields  
on a Ricci-flat background in \ci{Tseytlin:2013jya}.
For example,    while $RR h_1 h_3$ term    can not   lead to UV  divergent $RR$ term, a 
   non-trivial  term like $R_{\mu\nu\lambda\rho} \nabla_\mu  h_\kappa   \nabla_\lambda   h_{\nu \rho \kappa}   $ 
may.
}
\fi 
  

\section{Conclusions}

In this  paper we addressed the question of covariant description of conformal  higher spin fields in a 
non-trivial background. The standard  definition of   the CHS action \rf{133}  gives an  expansion 
near flat space  and thus is not   generally covariant.  Given that the spin 2 CHS   field   should have a natural interpretation 
of a conformal graviton,  one expects that there should be a  possibility to rewrite this action in a manifestly covariant form 
with the spin 2 part represented  by the  non-linear Weyl action.\foot{It should  be noted
that a possibility to rewrite the action for an infinite set of fields  in a manifestly covariant and local 
way  is not a priori obvious. For a  somewhat  related discussion in the string theory context see \ci{Tseytlin:1986eq}.} 

 We suggested a way  to define the CHS action   in a covariant way
by using  the background metric to  define the 
star product  in the associated particle  dynamics   and  thus in the definition of the  gauge transformations. 

As is well known,  the quadratic term  in an action expanded near its classical solution should have  linearized
gauge invariance. For  example,  the quadratic 4-derivative operator in the Weyl action expanded near Bach-flat background 
is consistent, i.e. 
has the standard reparametrization  invariance (which  is fixed  by a background gauge in quantum computations). 
The same was previously  found  to be true to linear order in the curvature expansion
 for the conformal spin 3 operator 
in a Bach-flat metric \ci{Nutma:2014pua}.
 Here we generalized  this  fact to any  conformal higher spin field
and commented  on a possibility of extending  this claim to terms quadratic in the curvature. 
We also pointed out the presence of  curvature-dependent mixing terms in the 
quadratic part of the conformal higher spin action  expanded  
in a non-trivial background. 



\section*{Acknowledegements}
We would like to thank  M. Beccaria, R. Metsaev,  R. Roiban, {E. Skvortsov} and M. Taronna   for useful discussions of related questions. 
{M.G. also wishes to thank N.~Boulanger for a useful discussion of Weyl invariants.}
AAT  is grateful to S. Kuzenko  for useful discussions and the hospitality during visit of  the University of Western Australia. 
This work was supported by the  Russian Science Foundation  grant 14-42-00047.
The work of AAT was also supported by the ERC Advanced grant No.290456, 
the STFC Consolidated grant  ST/L00044X/1  and  by the Australian Research Council,  project No. DP140103925.

\iffa 
It is useful to write  the  transformation of $K_s[g]$ under the Weyl transformation of the metric in \rf{3.7}   as  
\begin{equation}\la{317}
\delta_{\omega_0} K_{\mu_1\ldots \mu_s}=  \omega_0 K_{\mu_1\ldots \mu_s} + \delta^\prime_{\omega_0}K_{\mu_1\ldots \mu_s} \ . 
\end{equation} 
{\bf why $K_{\mu_1\ldots \mu_s}$ has Weyl weight 1 ?! }

Consider the variation of $\sqrt{g} K_s h^s$ under the deformed Weyl transformation. The only term proportional to $h_0$
is $\delta^\prime K^0 h_0$ and hence $\delta^\prime K^0=0$. This implies that $K^0$ is Weyl invariant with weight $2$ (i.e. behaves like a metric). Using~\rf{count} for $w=0$, for $s=0$ we get  $n_C=1$ and hence $K^0=0$ because $C$ is totally traceless.

Next, the terms in the variation proportional to $h_1$ vanish separately and hence
\begin{equation}
 K^0 (\delta^\prime h_0)|_1 +(\delta^\prime K^1_\mu )h^\mu_1=0
\end{equation}
where $(\delta^\prime h_0)|_1$ denotes the component proportional to $h_1^\mu$ in $\delta^\prime h_0)$. 
Because $K^0=0$ it follows $\delta^\prime K^1=0$ so that $K^1_\mu$ is also a conformal invariant of weight $2$. In this case equation~\rf{count} gives $n_c=1$ and $n_D=0$ however again at least two indices of $C$ have to be contracted by $g^{\mu\nu}$ and hence $K^1=0$ as well. 

In the same way at the next step one finds $\delta^\prime K^2=0$. In this case~\rf{count} has two meaningfull solutions
$n_C=1,\n_\cD=2$ and $n_C=2,\n_\cD=0$. The first one gives $\cD^\alpha \cD^\beta C_{\alpha\mu \beta \nu}$ as other contractions are either equivalent or give zero. This is proportional to Bach tensor and hence vanishes by assumption. The second one results in
\begin{equation} la{319} 
 \alpha g_{\mu\nu}C^2+\beta C_{\mu\alpha\beta\gamma} C_{\nu}{}^{\alpha\beta\gamma}\,.
\end{equation} 
This does not in general vanish on Bach flat  background.

Suppose we succeeded to show that $K^r=0$ for $r<s$. Then analogous arguments show that $\delta^\prime_{\omega_0} K_{\mu_1\ldots \mu_s}[g]=0$ and hence $K^s$ is a Weyl invariant of weight $w$.  This doesn't vanish in general. However, let us restrict to the analysis to linear order in the curvatures. At the zeroth order in $C$ one only has $g$ and as we are going to see later this vanish because $h$ can be assumed traceless. To first order $n_C=1, n_\cD=s$ and only 2 indices of the derivatives are contracted. Modulo higher order terms this again gives (derivatives of) the Bach tensor (see e.g.~\cite{Boulanger:2004eh}). So the present analysis only allows to put to zero linear in $h$ terms up to the second and higher order in curvatures.

Let us now analyze the structure of the linear in $h$ term in $S[g,h]$ from~\rf{310}. Because the action is invariant under diffeomorphisms it follows that $K_{\mu_1\ldots\mu_s}[g]$ is,  in fact, 
 a function of  only the 
 metric and  its  ``curvatures" (i.e. Riemann curvature and its covariant derivatives).
  Expanding $K$ in homogeneity in curvatures we can assume that the zeroth degree term is zero because it can only be a trace of $h_s$. To first order in the curvatures, $K$ has the following structure
\begin{equation}
 \nabla_{(\mu_1}\ldots \nabla_{\mu_l}R_{\nu\rho)}
\end{equation} 
where $R_{\nu\rho}$ denotes the Ricci tensor and where some of the indices (depending on the rank) are contracted by the metric.  Indeed, the generic expression is linear in covariant derivatives of the Riemann tensor and its upper index is contracted and some of the lower indices are contracted by the undifferentiated metric. Bianchi identities then imply that this amounts to terms of the above structure.

Furthermore, $S[g,h]$ is invariant under the following transformation
%
\begin{equation}
\label{w-const}
 \delta g^{\mu\nu}=\alpha^2 g^{\mu\nu}\,, \qquad \delta h^{\mu_1 \ldots \mu_s}=\alpha^2 h^{\mu_1 \ldots \mu_s}
 \end{equation} 
 where $\alpha=const$. Indeed, it is easy to check that the action
\rf{3.4} is invariant,  provided that $\phi$ transforms accordingly. The only subtlety is to check that  the symbol map does not contribute. This, in turn, follows from the fact 
 that the connection, the curvature $R_{\mu\nu\rho}^\sigma$  and  its covariant derivatives are invariant (i.e. have weight zero) under such transformation.
 At the same time,  $\sqrt{g}$ has weight $-d$ while the entire integrand should have vanishing weight.

Moreover, let us analyze the invariance of the linear in $h$ term under the transformation~\rf{39}. T this end let us restrict ourselves
to field configurations where all $h_t$ with $t>s$ vanish. The transformation~\rf{39} preserves such configurations because
one can not make $h_{t}$ nonvanishing if $t>s$. It follows that $\sum_{l=0}^s K_{\mu_1\ldots \mu_l} h^{\mu_1\ldots \mu_l}$ is invariant 
for such configurations. In particular, $K_{\mu_1\ldots \mu_s} h^{\mu_1\ldots \mu_s}$ has to be invariant on its own
 
 We now restrict to $d=4$. In this case $K$ should have weight 2 and hence contain just one undifferentiated metric. This gives precisely the terms showing up in the  Bach tensor \rf{55} in the linear in curvatures order. It is natural to expect that diffeomorphism invariance along with~\rf{w-const} implies that only the Bach tensor and its covariant derivatives may appear in $K$. 
 Similar arguments show that $S^0[g]$ can only be build  out of  Weyl tensors and hence in $d=4$ also leads to equations of motion proportional to the Bach tensor. 
 As we have already seen,  this implies that any Bach-flat $g_0$ leads to an exact solution  of the entire nonlinear theory (to first order in curvatures).  This in turn implies that quadratic in $h_s$ term in $S^2$ is gauge invariant (to first order in curvatures) on a  Bach-flat background.

%
\fi
\appendix
\section{Covariant quantization in 
Fedosov-type approach:\\
 quantum version of normal coordinate  expansion}
\label{sec:fedosov}

Let us recall how to perform quantization on the cotangent bundle in generic coordinates. Let $x^\mu$ be coordinates on the base manifold
and $p_\mu$ their conjugate momenta. The canonical Poisson bracket reads as $\pb{x^\mu}{p_\nu}=\delta^\mu_\nu$. We would like 
to define  quantization compatible with a given Riemanian metric $g_{\mu\nu}(x)$. Let us  introduce 
frame field $e^a_\mu$ and Lorentz connection $\omega_{\mu a}^b$ such that~\foot{We use convention $\nabla (T^ap_a)=dx^\mu \omega_{\mu b}^a T^b p_a$, $R_{\mu\nu\, b}^a=\d_\mu\omega^a_{\nu b}+\omega^a_{\mu c}\omega^c_{\nu b}-(\mu \rightleftarrows \nu)$. In particular, $\nabla^2f(x,y,p)=\qcommut{\half dx^\mu dx^\nu R_{\mu\nu\,b}^a y^b p_a}{f(x,y,p)}$.}
\begin{equation}
 \nabla e^a=0\,, \qquad \omega_a^b\eta_{bc}+\omega_c^b\eta_{ba}=0\,,\qquad g_{\mu\nu}=e^a_\mu e^b_\nu \eta_{ab}\,.
\end{equation} 
In what follows we will use the coordinates $x^\mu,\ p_a=e^\mu_a p_\mu$ on the cotangent bundle.

Let us introduce extra variables $y^a$ {which are coordinates on the tangent spaces} and the star product
\begin{equation}
 \circ=\exp\Big[\frac{\hbar}{2}(\dr{y^a}\dl{p_b}-\dr{p_a}{\dl{y^a}})\Big]\,.
\end{equation} 
\begin{prop}
Given $e,\omega$ there exist a nonlinear connection whose covariant derivative (acting on forms with values in functions of $y,p$)
has the form
\begin{equation}
D=dx^\mu\dl{x^\mu}+\hbar^{-1}\commut{e^a p_a+\omega_{\mu a}^b y^a p_b}{\ \cdot}_\circ+\hbar^{-1}\commut{\mathbf{r}}{\ \cdot}_\circ\,, \quad \mathbf{r}=
y^a y^b dx^\mu \mathbf{r}_{\mu ab}(x,y,p)
\end{equation} 
and for any $f(x,y,p)$ satisfies 
\begin{equation}
 DDf(x,y,p)=0 \ .
\end{equation} 
Under the  extra condition $e_a^\mu y^a\dl{(dx^\mu)}\mathbf{r}=0$ this  connection is unique and is such that $\mathbf{r}$ is linear in $p_a$.
\end{prop}
\begin{proof}
The proof is based on using suitable degree of homogeneity in $y$ plus and  $\hbar$ and acyclicity of the differential
$\delta=dx^\mu e_\mu^a \dl{y^a}$ in nonzero form-degree. 
\end{proof}

Note that {by construction} $D$ differentiates $\circ$-product. If $\mathbf{r}$ is linear in $p_a$ it also satisfies the Poisson bracket version of the
flatness condition (i.e. coincides with its classical limit). In what follows we assume that $D$ is minimal (satisfies $e_a^\mu y^a\dl{(dx^\mu)}\mathbf{r}=0$). Terms of degree 4 and less read explicitly as:\foot{We shall denote by $(...)$ 
some numerical coefficients   precise   values of which is not relevant for our discussion.}
\begin{multline}
 \te \mathbf{r}= dx^\mu \big[-\frac{1}{3}R^a_{\mu c b} p_a y^c y^b+(...)\nabla_d R^a_{\mu c b} p_a y^c y^b y^d+(...)\nabla_e\nabla_d R^a_{\mu c b} p_a y^c y^b y^d y^e+\ldots\big]\\
 +dx^\mu(R^c_{\mu d e} R^a_{f b c} p_a y^f y^b y^d y^e+\ldots)+\ldots\,.
\end{multline}
Here the first line contain terms linear in curvature and its covariant derivatives.\foot{Note that expansion in homogeneity in curvatures
$\mathbf{r}=\sum_{i=1}^\infty \mathbf{r}_i$ is well defined and the flatness condition decomposes as
\begin{equation}\no \te 
\nabla \r_i-\delta \r_i+\half\sum_{l+k=i,\,\,l,k>0}\commut{\r_l}{\r_k}_\circ=0
\end{equation} 
}
\begin{prop}\label{f-lift}
For any $f(x,p)$ there exist a unique $\check f(x,y,p)$ such that 
\begin{equation}
 D \check f=0\,, \qquad \qquad \check f|_{y=0}=f\,.
\end{equation} 
Moreover, if $D$ is a unique connection such that $e_a^\mu y^a\dl{(dx^\mu)}\mathbf{r}=0$
then for $f(x,-p)=\pm f(x,p)$  the associated $\check f$ also satisfies 
$\check f(x,-p)=\pm \check f(x,p)$. More precisely, for $f$ of homogeniety $s$ in $p_a$, $\check f$ contains terms
of homogeneity $s,s-2,s-4,\ldots$.
\end{prop}
\begin{proof}
$\check f$ is constructed iteratively in  the degree of homogeneity in $y$ and $\hbar$.
For $D$ special $\mathbf{r}$ is linear in $p_a$ so that the star commutator may only reduce the homogeneity in $p_a$ by an even number.
\end{proof}
It follows that the space of all functions in $x,p$ is isomorphic to covariantly constant functions depending in addition on $y$-variables.
{Below and in the main text} we need the following example:  if  $\eta=\half \eta^{ab}p_a p_b$ then 
\begin{equation}\te 
\check\eta=\half \eta^{ab}p_ap_b+\frac{1}{6}R^a_{bcd}p_a p^b y^c y^d+ (\text{terms of degree $>2$ })
\end{equation} 
This  is related to the expansion in normal coordinates if one identifies $y^a$ as normal coordinates around $x^\mu$. Note that, in general, terms independent of momenta may appear but they are of order $\hbar^2$. For a general element $f=f^{ab}(x)\, p_ap_b$  quadratic in $p_a$ one has
\begin{multline}
 \check f=f^{ab}(x)p_ap_b+y^a\nabla_a f^{bc}p_bp_c+\half y^a y^b \nabla_a\nabla_b f^{cd}p_cp_d
 +(...)R^a_{bcd} f^{be} y^c y^d p_a p_e+\\ \te +\frac{2}{3} \hbar^2 R^a_{bcd}\nabla_a f^{cd} y^b+\ldots
\end{multline} 
where dots denote terms of total degree higher than $2$. For  a linear one
\begin{equation}
\label{linplift}
 \check f=f^a p_a+y^b\nabla_b f^a p_a+\half y^a y^b \nabla_a\nabla_b f^{c}p_c
 +(...)R^a_{bcd}f^by^cy^dp_a+\text{(deg $\geq 3$ terms)}
\end{equation} 

The above construction gives the covariant $*$ product on contangent bundle:  using the above propositions  we may define 
\begin{equation}
 f*g:=(\check f\circ \check g)\big|_{y=0}\,.
\end{equation} 
The consistency of this definition 
follows from the fact that for any $\check f, \check g$ satisfying $D \check f=D \check g=0$ one has
$D(\check f\circ \check g)=0$.  
{The above construction of the star product is a version of that of~\cite{Bordemann:1997er} which in turn has its 
origin in the Fedosov quantization~\cite{Fedosov:1994}}.
% ... 

As an example let us compute explicitly the tranformation of the spin $1$ under the tranformation generated by $\epsilon^{ab}p_ap_b$:
\begin{equation}
\label{gs-3-1}\te 
\delta (h^{a}p_a)=\commut{-\half\eta^{ab}p_a p_b}{\epsilon^{cd}p_cp_d}_*{}\big|_{1}=-\frac{4}{3}R^a_{bcd}\nabla_a\epsilon^{cd}p^b\ . 
\end{equation}

Next, let us  describe the representation space in a covariant way. 
Let $\rho$ denote a map that sends Weyl symbol $f(y,p)$
into the respective operator in coordinate  representation (i.e.  on functions of $y$). For instance,  
$\rho(y_a p_b)=-\half \hbar(y_a\dl{y_b}+\dl{y_b}y_a)$. 
\begin{prop}\label{phi-lift}
For any wave function $\phi(x)$ there exist a unique lift $\check\phi(x,y)$
satisfying\\
 $\hbar\left[\nabla+\rho(e^a p_a+\r)\right]\check\phi=0$ and $\phi|_{y=0}=\phi$.
\end{prop}
To illustrate this, let us explicitly evaluate the lift up to terms of degree $3$:  
\begin{equation}
 \check\phi=\phi+y^a \nabla_a \phi+\half y^a y^b\big[\nabla_a \nabla_b + (...)\hbar^2 R_{ab}\big]\phi+\ldots
\end{equation} 
The action of the operator $\hat f(x,\dl{x})$ with symbol $f(x,p)$ on the wave function $\phi(x)$ is defined by
\begin{equation}
\label{symb-map}
 \hat f \phi=\left(\rho(\check f)\Phi\right)|_{y=0}\,.
\end{equation} 
Note that by construction $1$ acts as an identity operator and $\hat{(f*g)}=\hat f \,\,\hat g$ (because $\rho$ is a representation map).
This way we have constructed a covariant symbol map that sends functions of $x,p$ to differential operators on $x$. Note that the map is solely expressed in terms of covariant derivatives, frame field, and curvature (along with its covariant derivatives). This shows that although the map is written in terms of local coordinates and local frame it does not depend on the choice of coordinates and the frame.

The above technique allows to reformulate the relations 
 \rf{gtr} and \rf{211} in  manifestly coordinate-independent terms. 
 In so doing the 
component fields entering $H({x,p})$ transform as tensors under a change of coordinates. 
 By a suitable field redefinition
one can also achieve that they transform homogeneously under the linearized gauge  transformations (see the end of  this appendix 
 for spin 2 case).

Let us now discuss the inner product. The minimal choice is 
\begin{equation}
 \inner{\phi}{\chi}=\int d^dx \sqrt g \,\, \phi^*(x)\, \chi(x) 
\end{equation} 
The question  is how to identify (anti)hermitian operators at the level of symbols. 
\begin{prop}
Real (imaginary) symbols correspond to hermitian (antihermitian) operators.
\end{prop}
\begin{proof}
First of all we show that for $f(x,p)$ real (imaginary) the respective lift $\check f(x,p,y)$ is also real (imaginary). Let us for definiteness consider real $f$. It is enough to assume all coefficients to be  real so that $f(x,p)$ contains only even powers of $p_a$.
By inspecting the recursive construction of $\check f$ we see that odd powers of $p$ can  not  appear as well as imaginary coefficients
(we assume that the  metric,  frame field and  connection  are real). Finally, because $\check f$ is real $\rho(\check f)$ is formally hermitian when represented  on wave functions of $y^a$ where the conjugation rules are 
$(y^a)^\dagger=y^a$ and $\dl{y^a}^\dagger=-\dl{y^a}$
(in this case dependence on $x^\mu$ is irrelevant and as before ${x^\mu}^\dagger=x^\mu$).

It is enough to check this  statement for operators whose symbols are of zeroth and first order in $p$.  Indeed, 
 such operators generate the entire algebra. For $f=f(x)$ the statement is obvious. For $f=v^a(x)p_a$ we have (this is just a rewriting of~\rf{linplift})
\begin{equation}
 \check f(x,p,y)=\tilde v^a(x,y)p_a\,, \qquad \tilde v^a=v^a+y^b\nabla_b v^a+O(y^2)
\end{equation}
Because $\rho(f)$ is formally antihermitian on wave functions of $y$  we have
\begin{align}
 \int d^dx \sqrt g \,\, \phi^* \hat f \chi &=\int d^dx \sqrt g \,\, \left(\check \phi^* \rho(\check  f) \check \chi\right)\big|_{y=0}\no \\
 &=\,\half\int d^dx \sqrt g \,\, \left(\check \phi^* (\tilde v^a \dl{y^a}+\dl{y^a}\tilde v^a) \check \chi\right)\big|_{y=0}\la{aa1} \\
 &=\,-\int d^dx \sqrt g \,\, \left((\rho(\check  f) \check \phi)^*  \check \chi\right)\big|_{y=0}+
 \int d^dx \sqrt g \,\, \left(\dl{y^a} (\check \phi^* \tilde v^a \check \chi\right))\big|_{y=0}\,. \no 
 \end{align}
Using that $(\dl{y^a}X)|_{y=0}=\nabla_a Y$ for some $Y$, where $X$ is $\phi^*$ or $\chi$ or $\tilde v^a$,  the integrand of the last term
can be rewritten as $\sqrt g \nabla_a X^a=\sqrt g \nabla_\mu X^\mu=\d_\mu(...)^\mu$ and hence the integral vanishes under the standard assumptions.
\end{proof}
To summarize, we have constructed a covariant (independent of the choice of local coordinates) description of quantum mechanics
 on the cotangent bundle. We  thus have all  the required ingredients:
 representation space, inner product, operators, symbols and symbol-map.

\section{Weyl invariants}
\label{sec:weyl}

Let us briefly recall some  known results on the structure of the conformal and diffeomorphism invariants. More precisely, we are interested in (tensor valued) local functions
$K_{\m_1...\m_s}$   of the metric and its derivatives  (cf. \rf{3100})  
that transform covariantly under the diffeomorphsims and Weyl transformations. It turns out that a candidate invariant is a polynomial in
\begin{equation}\la{314}
 g_{\mu\nu},~~ C_{\mu\nu\rho\sigma},~~\cD_\alpha C_{\mu\nu\rho\sigma}, ~~\cD_\alpha \cD_\beta C_{\mu\nu\rho\sigma},~~\ldots
\end{equation} 
with  indices  properly contracted by $g^{\mu\nu}$. Here $C_{\mu\nu\rho\sigma}$ is the  Weyl tensor 
and $\cD_\alpha$ denotes a Weyl-covariant derivative related to the so-called Thomas D-derivative.\foot{The first $\cD$-derivative of Weyl tensor is the same as the ordinary covariant derivative , i.e. 
$\cD_\alpha  C_{\beta\gamma\delta\rho} = \nabla_\alpha  C_{\beta\gamma\delta\rho} $.}
In general, such  polynomial is not  invariant  under Weyl transformations 
as in contrast to $g^{\mu\nu}$ and $C_{\mu\nu\rho\sigma}$ the transformation law of $\cD_\alpha \ldots \cD_\beta  C_{\mu\nu\rho\sigma}$  may  involve  a gradient of the Weyl parameter $\omega_0$. Hence the invariance condition imposes extra  constraints on the structure of the polynomial.
For more details we refer to~\cite{Boulanger:2004eh} and references  there.

Let us analyze  the necessary condition for a rank $s$ tensor-valued local function to be diffeomorphism covariant and Weyl covariant with weight $w$. Taking into account that $g^{\mu\nu}$ has Weyl weight $-2$ while $g_{\mu\nu}$ and $\cD_\m{\ldots}\cD_\n C_{\alpha\beta\gamma\delta}$ have weight $2$, we get
\begin{equation}\la{315}
 2n_g+4n_C+n_\cD-2n^g=s\,, \qquad \qquad -2n_g-2n_C+2n^g=-w\,,
\end{equation} 
where $n_g,\, n_	C,\, n_\cD$ and $ n^g$ denote, respectively,  the numbers of $g_{\mu\nu},C,\cD$ and $g^{\mu\nu}$  factors in a polynomial.  
 The first equation counts indices while the second counts the Weyl weight. As a consequence,  we have
\begin{equation}
\label{count}
2n_C+n_\cD=-w+s \ . 
\end{equation} 
Consider, for example,  a scalar invariant which is an integrand of 
$S_0[g]=\int d^4x \sqrt{g} L_0$. One finds that $L_0$ is Weyl invariant of weight $-4$ for which~\rf{count} has two solutions
$n_C=1,n_\cD=2$ and $n_C=2,n_\cD=0$. The first one gives zero (as $C$ is traceless)
 so 
one ends up with  $L_0=C^2$, i.e. the well-known   Weyl   gravity  Lagrangian.

Next, let us  consider  a 
rank-one tensor $K_\mu$ of weight $w=-2$ appearing in \rf{3100}.
Then we have only one nontrivial solution: $n_C=1$,
$n_\cD=1$.  It should  again  vanish as here  at least two indices of
 the Weyl tensor   should  be contracted with the metric. 

For a polynomial  $K_{\m\n}$  with $s=2,w=-2$ we have two solutions:
 $n_C=2$, $n_\cD=0$ and $n_C=1$, $n_\cD=2$. The latter one
necessarily contains two derivatives contracted with the indices of $C$ and hence should vanish on a Bach-flat background. The former  can be brought to the following form
\begin{equation}\la{b4} 
k_1   g_{\mu\nu}C_{\alpha\beta\gamma\rho}  C^{\alpha\beta\gamma\rho} +  k_2 C_{\mu\alpha\beta\gamma}C_{\nu}{}^{\alpha\beta\gamma}\,.
\end{equation} 
Imposing the  tracelessness ($k_1 = - {1\ov 4} k_2$)  and covariant conservation conditions~\eqref{covcons}
this can be shown to vanish on a Ricci-flat background using $ \nabla_{[\m} C_{\n\alpha]\beta\gamma}=0$; Weyl-covariance implies that same should be true on a  Bach-flat background. 

\iffa 
In the  $s=3,w=2$ case  we get the following options: 
$n_C=1$, $n_\cD=3$ and $n_C=2,n_\cD=1$. The first one is proportional to derivatives of the Bach tensor  and thus   vanishes on a  Bach-flat background. 
 The second one  involves 
\begin{equation}
  C_{\mu\alpha\beta\gamma}\nabla_\nu C_{\rho\sigma\delta\upsilon}
\end{equation} 
with  3 pairs of indices  contracted by the metric $g^{\m\n}$.   Possible nontrivial contractions are
\begin{equation}
 k_1 C_{\mu\alpha\beta\gamma}\nabla_\nu C_{\rho}{}^{\alpha \beta \gamma}+ k_2 
  C_{\mu\alpha\beta\gamma}\nabla^\beta C_{\nu}{}^{\alpha}{}_\rho {}^\gamma+
  k_3 C_{\mu\alpha\nu\gamma}\nabla^\beta C_{\beta}{}^{\alpha}{}_\rho {}^\gamma
\end{equation} 
where the symmetriazation of $(\mu\nu\rho)$ is assumed.
In addition it needs to satisfy~\eqref{covcons}. 
{\bf It is not clear if  this implies  that 
it vanishes on a  Bach-flat  background}.
\fi

The analysis for $s>2$ becomes rather involved. Considerable simplification can be achieved by employing the spinor formalism
in $4d$. 
In this approach the self-dual (anti-self-dual) component  of $C$ is represented 
 by the totally symmetric spinor $\Psi_{ABCD}$
($\Psi_{A'B'C'D'}$) where  $A=1,2$ ($A'=1,2$). The invariant contractions of indices are performed with the help of the antisymmetric tensor $\epsilon^{AB}$  or $\epsilon^{A'B'}$. In particular, the  Minkowski metric $\eta_{\mu\nu}$ in spinorial notations reads  as  $\eta_{AA', BB'}=2\epsilon_{AB}\epsilon_{A'B'}$
(for a concise exposition  see, e.g.,~\cite{Didenko:2014dwa}  and refs. therein). 

For example, for $s=2$, by writing the spinorial counterpart of~\eqref{b4} one finds that the second term necessarily vanishes  so that the tracelessness condition
implies that the first term vanishes as well.

For $s=3$ we have $K_{\mu\nu\rho}$ which,  according to~\eqref{count},  can not have terms of order higher than 2 in Weyl tensor and its Weyl-covariant derivatives. As the linear in $C_{\mu\nu\rho\sigma}$ term vanishes on Bach-flat background let us concentrate on the quadratic contribution which   should 
involve only one covariant derivative. In the spinorial  approach  $K_{\mu\nu\rho}$
is described by  $K_{AA'BB'CC'}$  to which only the following terms
may contribute 
\begin{equation}
 \Psi_{ABCD} \nabla_{EE'}\Psi_{A'B'C'D'}, \qquad 
\Psi_{A'B'C'D'}\nabla_{EE'}\Psi_{ABCD}\ , 
\end{equation} 
where the indices are contracted with  the $\epsilon$-tensors. It is clear that there is only one inequivalent contraction that leaves $3+3$
free indices. It results in the following general expression:
\begin{equation}
 n_1\,\Psi_{ABCD} \nabla^{DD'}\Psi_{A'B'C'D'}+
n_2\,\Psi_{A'B'C'D'}\nabla^{DD'}\Psi_{ABCD}\,.
\end{equation} 
The Weyl covariance of $K_{\mu\nu\rho}$ implies that $n_2=-n_1$ in which case the above expression is
proportional to 
 the Eastwood-Dighton tensor $E_{\mu\nu\rho}$  in   \rf{ed}. It  is known to vanish for the 
 metric conformal to the Einstein one.
 Note that the Eastwood-Dighton tensor is by construction trace-free and its divergence 
 is proportional to the Bach tensor    and thus 
 vanishes on a Bach-flat background.
 
{Let us  note  that all $K_s$ vanish in the special case of Bach-flat 4d backgrounds with self dual  (or antiselfdual) Weyl tensor. In this case 
 $\Psi_{A'B'C'D'}=0$ so that $K_s$ is build out of $\Psi_{ABCD}=0$, $\epsilon_{AB}$, $\epsilon_{A'B'}$, $\cD_{AA'}$. Moreover, $\epsilon$ may only enter to contract indices because $K_s$ should be totally symmetric so that primed indices may only originate from covariant derivatives so that \eqref{count} implies $n_C=1$ and hence $K_s$ should be proportional to the Bach tensor.}

. 
Indeed, in this case $K_3$ and $K_4$  vanish  just on the basis of the index structure.

  
Finally, let us list two  useful  relations   in spinorial notations. The Bianchi identity for
the  Weyl tensor reads
\begin{equation}
\nabla^A_{B'}\Psi_{ABCD}=\nabla^{A'}_{B}\Phi_{CDA'B'}-2\epsilon_{B(C}\nabla_{D)B'}\Lambda\,,
\end{equation}
where $\Phi_{ABA'B'}$ is the  trace-free Ricci spinor and $\Lambda$  is a multiple of the scalar curvature. The Bach tensor  is given by 
\begin{equation}
 B_{AA'BB'}=2(\nabla^C{}_{A'}\nabla^D{}_{B'}+\Phi^{CD}{}_{A'B'})\Psi_{ABCD}=2(\nabla^{C'}{}_{A}\nabla^{D'}{}_{B}+\Phi^{C'D'}{}_{AB})\Psi_{A'B'C'D'}\,.
\end{equation}


\newpage

\iffa 

\section{Ambient Fedosov-type approach}

In terms of the ambient space (in the fiber) the CHS fields on conformally-flat space can be described by the following equations
\begin{equation}
 \begin{gathered}
\nabla \Phi=0\,,\qquad  \d_Y\cdot \d_Y \Phi=\d_Y\cdot \d_P \Phi=\d_P\cdot \d_P \Phi=0\\
 (Y+V)\cdot \d_{Y}\Phi=(P\cdot\d_{P}-(Y+V)\cdot\d_{Y}-2)\Phi=0\\
\end{gathered}
\end{equation} 
along with gauge equivalence:
\begin{equation}
 \begin{gathered}
\Phi \sim \Phi+P\cdot \dl{Y}\chi
\end{gathered}
\end{equation} 
where $\chi$ is a gauge parameter satisfying exactly the same constraints as $\Phi$ except that the last one is replaced by 
$(Y+V)\cdot \dl{Y}\Phi=(P\cdot\dl{P}-(Y+V)\cdot\dl{Y})\Phi=0$.
Here $\Phi,\chi$ are fields taking values in formal series in $Y^A$ and polynomials in $P_A$ (these variables contract tensor indices).
$\nabla$ denotes $o(d,2)$ covariant derivative in the specific representation:
\begin{equation}
 \nabla=dx^\mu(\dl{x^\mu}-\cA_{\mu A}^B\left[ (Y^A+V^A)\dl{Y^B}+P^A\dl{P_B}\right])
\end{equation} 
where $\cA_{\mu A}^B$ is the usual $o(d,2)$ flat connection which in the standard basis $+,a,-$ is typically taken in the form
\begin{equation}
\label{conn-comp}
\cA_-^a=e^a\,, \qquad \cA^a_b=\omega^a_b\,, \qquad \cA^a_+=\Rho^a\,.
\end{equation}
It is clear that $e^a=dx^\mu e_\mu^a$ is identified with the frame field associated to $g_{\mu\nu}$, $\omega^a_b$ with the spin connection
and $\Rho_b=dx^\mu e_\mu^c \Rho_{cb}$ is related to a Rho-tensor $\Rho_{cb}$ (trace adjusted Ricci tensor) of $g_{\mu\nu}$.
The remaining nonvanishing components are determined by antisymmetry of $\cA^{AB}$
\begin{equation}
 \cA_b^+=-e_b\,, \qquad \cA^-_b=-\Rho_ b\,.
\end{equation} 
Furthermore, $\cA^+_-=\cA^-_+=0$. In this form $\cA$ is known as a tractor connection and $\cT(M)$ as a tractor bundle. The curvature
of $\cA$ contains the Weyl tensor $d\omega^a_b+\omega^a_c\omega^c_b - \Rho^a e_b-\Rho_b e^a$ and the torsion $de+\omega e$
associate to metric $g_{\mu\nu}$ which vanish in the conformally flat case.


\subsection{Factorization of the scalar kinetic oprator}
Consider the simplest case of scalar. In the conformally flat case the (higher-order) conformal scalar equations can be seen as arising
from the critical scalar in the bulk. This is described by:
\begin{equation}
 D\Phi(x,Y)=0\,, \qquad \Box\Phi=0\,, ((Y+V)\cdot \dl{Y}+\Delta)\Phi=0\,.
\end{equation} 
More precisely, considering these equations on the conformal space described by connection $\cA$ and compensator $V, V^2=0$ it turns out that 
for $\Delta=\frac{d}{2}-\ell$ with $\ell\in \mathbb{N}$ these equations encode conformal powers of Laplacian, e.g. in flat space simply 
$\eta^{\mu\nu}\dl{x^\mu}\dl{x^\nu}\phi(x)=0$, where $\phi(x)=\Phi(x,Y)|_{Y=0}$.

To see how it works in generic conformally flat background let us explicitly write down 
\begin{equation}
 \nabla_a=e^\mu_a \nabla^L_\mu+
\end{equation}

\fi

\iffa 

 actually, I   now   do not understand what   is said in that foot 3  in my paper -- 
let me review the general   point about 3 and higher vertices: 
we assume  that linearized action  is 
\begin{equation}
\int d^4 x C_s C_s 
\end{equation}
with sum over $s=0, 1,2,...$,  $C_s = \d^s \phi_s$,  $\phi_s$ has dim $2-s$    so $C_s$  has dim 2.
Interacting theory must not have dim parameters so 
local cubic vertex may only be 
\begin{equation}
\int d^4 x  \d^{s_1+s_2 + s_3 -2}   \phi_{s_1} \phi_{s_2} \phi_{s_3}                                                          
\end{equation} 
this  could  in theory  be written as    
\begin{equation}
\frac{1}{\d^2}  C_{s_1} C_{s_2} C_{s_3}  
\end{equation} 
but that will be nonlocal.
 local action  between one spin 2   field  and 2 others can be 
\begin{equation}
 \int d^4 x \, \d^{s_1+s_2 }   \phi_{s_1} \phi_{s_2}   \phi_2 
\end{equation} 
and to make sense   that   can  only have $s_1=s_2$   so 
 this is naturally 
\begin{equation}
\int d^4 x    \phi_2   C_s  C_s 
\end{equation} i.e.    say contraction of metric with   2 linearized spin s tensors or covariantization of 
linearized $C_s  C_s$ 
if we have  2   spin 2   fields in the vertex 
\begin{equation}
\int d^4 x  \d^{s + 2 }   \phi_s \phi_2   \phi_2 
\end{equation} 
that may  make sense only for $s<3$   to be able to contract indices ?
but we may put many derivatives on  $\phi_2$. 
If $\phi_2$ satisfies its eqs of motion all such  cubic vertices should be consistent with spin s gauge invariance. 
general cubic relations should be Bekaert, Joung, Mourad
explicit spin 3 case is in Nutma-Taronna
http://arxiv.org/pdf/1404.7452.pdf


\fi

\iffa 
\subsection{Covariant expansion of the induced CHS action}
we start with Segal action
$S(h_2,h_s) =  \text{$log$-divergen part of~~}\tr \log ( \Box  +  h_s   \d^s)$
and assume that there is such redefinition of $h_s$ that puts it into
covariant form wrt  the metric
\begin{equation}
S'(g, H_s)=  \tr \log (  - D^2 + R/6   +  H_s    D^s)
\end{equation} 
where $H_s$  are  redefined higher fields  and D are now covariant derivatives.

Segal's   symmetries then should take   the form of  usual  one for $g_{mn}$
and  some deformed symmetries for $H_s$.

now  let us expand  $S$ to quadratic  order in $H_s$:
\begin{equation}
S=    C^2  +   H_3  ( D^6 + ... ) H_3  +     H_4 ( D^8 +  ...) H_4
 +    \sum c_{pqr}  H_p H_q H_r   + ....
\end{equation} 
This action by construction has right  flat limit and linearised
symmetries.

assumption:   there are no mixed  terms like
$D^k H_3 D^n H_4  C^p$   etc   so all is diagonal in $H$  at quadratic in $H$ order.

This is equivalent to assumption that  consistent vac  solutions are
$\text{Bach}=0,H_s=0$.
Note that this is not so in Vasiliev's theory: $R_{mn} = L g_{mn}$ is not a
consistent solution (apart from trivial AdS or dS)      if all other
fields $=0$.

Then next assumption is that  under gauge transformation of $H_3$
$\delta H_3 = D e_3  + ...$, $\delta   g_{mn} =    e_3  D H_3  + .... $
or smth similar allowed  by dimensions --
$[g]=0, [H_s] = 2 -s ,    [e_s] = 1-s$
$\delta H_4  =  \text{whatever}$.

Then  to quadratic order in $H_s$  we will have under $e_3$ transf
$C^2 -- >  \text{Bach}   e_3  D H_3  + ....
 H_3  ( D^6 + ... ) H_3   -- >  H_3  ( D^6 + ... )  D   e_3   + ...
H4 ... H4 --- >  H4  ( ...)  (...)$

Thus if whole action is invariant,  then   for $\text{Bach}=0$ background
$H_3 H_3$ term must have usual $\delta H_3 = D e_3$  invariance.

This was explicitly shown by Taronna-Nutma  by brute force.
So their result supports this picture.
\fi


\bibliography{CHS-Biblio2}
\end{document}